\documentclass[preprint2]{aastex}

\shorttitle{Inner Galaxy \ion{H}{2} Regions}
\shortauthors{Simpson et al.}

\newcommand{\teff}{$T_{\rm eff}$} 
\begin{document}

\title{On the Measurement of Elemental Abundance Ratios in Inner Galaxy \ion{H}{2} Regions}

\author{Janet P. Simpson\altaffilmark{1}, Robert H. Rubin\altaffilmark{2}, Sean W. J. Colgan, Edwin F. Erickson, and Michael R. Haas}
\email{simpson@cygnus.arc.nasa.gov; rubin@cygnus.arc.nasa.gov; Sean.Colgan@nasa.gov; erickson@cygnus.arc.nasa.gov; Michael.R.Haas@nasa.gov}

\affil{NASA Ames Research Center}
\affil{MS 245-6, Moffett Field, CA 94035-1000}
\altaffiltext{1}{SETI Institute}
\altaffiltext{2}{Orion Enterprises}

\begin{abstract}

Although variations in elemental abundance ratios in the Milky Way Galaxy certainly exist, 
details remain uncertain, particularly in the inner Galaxy, where stars and \ion{H}{2} regions 
in the Galactic plane are obscured optically.
In this paper we revisit two previously studied, inner Galaxy \ion{H}{2} regions:
G333.6$-$0.2 and W43.
We observed three new positions in G333.6$-$0.2 with the Kuiper Airborne Observatory
and reobserved the central position with the Infrared Space Observatory's 
Long Wavelength Spectrometer in far-infrared lines 
of S$^{++}$, N$^{++}$, N$^{+}$, and O$^{++}$. 
We also added the N$^+$ lines at 122 and 205 \micron\ 
to the suite of lines measured in W43 by Simpson et al. (1995).
The measured electron densities range from $\sim 40$ to over 4000 cm$^{-3}$ 
in a single \ion{H}{2} region,
indicating that abundance analyses must consider density variations,
since the critical densities of the observed lines range from 40 to 9000 cm$^{-3}$.
We propose a method to handle density variations and 
make new estimates of the S/H and N/H abundance ratios.
We find that our sulfur abundance estimates for G333.6$-$0.2 and W43 agree with the 
S/H abundance ratios expected for the S/H abundance gradient previously reported by Simpson et al., 
with the S/H values revised to be smaller owing to changes in collisional excitation cross sections.
The estimated N/H, S/H, and N/S ratios are the most reliable because of their small corrections 
for unseen ionization states ($\lesssim 10$\%).
The estimated N/S ratios for the two sources are smaller than 
what would be calculated from the N/H and S/H ratios in our previous paper. 
We compute models of the two \ion{H}{2} regions to estimate corrections 
for the other unseen ionization states.
We find, with large uncertainties, that oxygen does not have a high abundance, 
with the result that the N/O ratio is as high ($\sim~0.35$) as previously reported.
The reasons for the uncertainty in the ionization corrections for oxygen are both the non-uniqueness
of the \ion{H}{2} region models 
and the sensitivity of these models to different input atomic data and stellar atmosphere models.
We discuss these predictions and conclude that only a few of the latest models adequately 
reproduce \ion{H}{2} region observations, 
including the well-known, relatively-large observed Ne$^{++}$/O$^{++}$ ratios 
in low- and moderate-excitation \ion{H}{2} regions.

\end{abstract}


\section{Introduction}

The presence of radial abundance gradients in the plane of the Milky Way 
is now an established fact and is seen in both stars and gaseous nebulae
(see, e.g., Henry \& Worthey 1999; Rolleston et al. 2000; and references in both papers).
The cause is generally thought to be due to the formation history 
and subsequent evolution of the Galaxy;
thus the observed gradients are a major tool for understanding this history
(see, e.g., Hou, Prantzos, \& Boissier 2000; Chiappini, Matteucci, \& Romano 2001; 
Chiappini, Romano, \& Matteucci 2003, and references therein).
The basic idea is that the inner Galaxy formed before the outer Galaxy
and the higher molecular gas density in the inner Galaxy produces a higher star formation rate.
The result is a greater return to the interstellar medium (ISM) in the inner Galaxy 
of both ``primary'' alpha elements from massive star supernovae 
and ``secondary'' elements like nitrogen. 
Secondary nitrogen is produced by CNO burning of already existing carbon and oxygen 
in intermediate-mass stars and subsequently returned to the ISM through mass loss.
However, both the wide range of uncertain input parameters to chemical evolution models 
(Pagel 2001) and the uncertain details of the abundance variation of each element,
primary and secondary, contribute to our less-than-complete understanding 
of the formation and evolution of the Milky Way.  
In this paper we try to improve our knowledge of the abundances in inner Galaxy 
\ion{H}{2} regions 
and also test the methods that have been used to date 
to determine the abundance ratios from low-excitation, optically obscured \ion{H}{2} regions
that are needed for abundance gradient calculations.

\ion{H}{2} regions in the inner Galaxy 
(galactocentric radius $R_G \lesssim 6$ kpc, where $R_\odot = 8$ kpc) 
are not accessible to optical observers because of interstellar extinction. 
Consequently, the abundances must be determined from far-infrared (FIR) lines,
such as those of N$^{+}$ at 122 and 205 \micron, N$^{++}$ at 57 \micron,
O$^{++}$ at 52 and 88 \micron, Ne$^{+}$ at 12.8 \micron, 
Ne$^{++}$ at 15.5 and 36 \micron, S$^{++}$ at 18.7 and 33.5 \micron, 
and S$^{3+}$ at 10.5 \micron.
Large surveys include those of 
Simpson et al. (1995a, hereafter SCREH) and Afflerbach et al. (1997), who 
observed the FIR lines (wavelengths $>$ 17 \micron) with the Kuiper Airborne Observatory (KAO)
in 18 inner Galaxy \ion{H}{2} regions, and 
Mart\'{\i}n-Hern\'{a}ndez et al. (2002a; 2003), who observed 13 inner Galaxy \ion{H}{2} regions with the 
Infrared Space Observatory (ISO).
The KAO observations included FIR measurements of the lines with wavelength $> 18$ \micron,
whereas the ISO observations included the mid-infrared lines, 
measured with ISO's Short Wavelength Spectrometer, SWS, but not 
the [\ion{N}{2}] 205 \micron\ line.

Two of the abundance gradients that are most needed 
for galactic chemical evolution studies 
are the O/H and N/H ratios, representing primary and secondary elements, respectively.
In optically obscured \ion{H}{2} regions the only ionization state of oxygen available for study
is O$^{++}$, which means that corrections are required for the unseen O$^+$ ions.
The situation is better for nitrogen, as there are [\ion{N}{2}] lines as well as [\ion{N}{3}] lines 
in the FIR wavelength range.
Unfortunately, 
the [\ion{N}{2}] 122 \micron\ line is difficult to measure for airborne observations 
because of atmospheric absorption,
and 
the [\ion{N}{2}] 205 \micron\ line is difficult because of poor detector sensitivity.
In addition, the [\ion{N}{2}] lines are difficult to interpret 
because they are collisionally de-excited at much lower densities
than the lines from the other ions.
Consequently,
the papers to date have used only the [\ion{N}{3}] and [\ion{O}{3}] lines to estimate N/O
and the secondary/primary abundance ratio in optically obscured \ion{H}{2} regions
(SCREH; Afflerbach et al. 1997; Mart\'{\i}n-Hern\'{a}ndez et al. 2002a).
This is particularly troubling since a majority of the nitrogen and an even larger fraction of the oxygen 
are singly ionized in the high-metallicity \ion{H}{2} regions of the inner Galaxy 
(note that in the outer Galaxy, the lower metallicity \ion{H}{2} regions 
can be much more highly excited and
thus the N$^{++}$/O$^{++}$ ratio is often a good approximation to N/O, Rudolph et al. 1997).
On the other hand, a majority of the primary element sulfur is doubly ionized in \ion{H}{2} regions.
Here we devise a method of analyzing the density-sensitive [\ion{N}{2}] lines to more 
accurately estimate nitrogen abundances with respect to sulfur without the need
for large corrections for unseen ionization states.
Using the observed N$^{++}$/N$^{+}$ ratio and calculated \ion{H}{2} region models,
we then estimate the ionization correction factors (ICFs) for oxygen.

Two of the most luminous \ion{H}{2} regions in the Galaxy are
G333.6$-$0.2 at $R_G \sim 5.5$ kpc (Wilson et al. 1970) or 5.6 kpc (Caswell \& Haynes 1987)
and W43 (G30.8$-$0.0) at $R_G \sim 4.2$ kpc (Reifenstein et al. 1970) for $R_\odot = 8.0$~kpc.
The radio fluxes from these objects require ionizing stellar luminosities 
of $10^{50}$ and $2.9\times10^{50}$ photons s$^{-1}$, respectively (see Rubin 1968b).
Both \ion{H}{2} regions are located in the Scutum-Crux spiral arm (see Vallee 2002)
in the Milky Way molecular ring.
Both are excited by clusters of massive stars, observed at near-infrared (NIR) wavelengths.
The cluster exciting G333.6$-$0.2 contains stars that are 
very young, embedded in dust (Blum et al. 2002), in keeping with the youthful appearance 
of this extremely dense, compact \ion{H}{2} region.
On the other hand, the cluster exciting W43 must
be older and more evolved because it contains a Wolf-Rayet star in addition to 
O giants or supergiants (Blum, Damineli, \& Conti 1999; Cotera \& Simpson 1997),
although there are still numerous stars forming in the associated molecular clouds
(Motte, Schilke, \& Lis 2003).
This is also consistent with the more mature appearance of W43, where the ionized gas is
well separated from the exciting star cluster.

In this study, we report measurements of the [\ion{N}{2}] lines
in the three positions observed in W43 by SCREH 
and numerous lines in three additional positions in G333.6$-$0.2 north of the
center position observed by Colgan et al. (1993) with the KAO;
we also report measurements of the central position in G333.6$-$0.2
with ISO's Long Wavelength Spectrometer (LWS).
In Section 2 we describe the observations, 
in Section 3 we discuss analysis techniques that are used to determine abundances 
in both low- and high-density \ion{H}{2} regions, 
in Section 4 we model the two \ion{H}{2} regions to 
determine the ICFs needed for those elements
without measured abundances of the singly ionized atoms, 
and in Section 5 we discuss models in general and give our conclusions.
By measuring both N$^{+}$ and N$^{++}$, we hope to avoid the need for large 
(and uncertain) ICFs for nitrogen and thus improve the reliability 
of the secondary/primary abundance ratio measurement in the 
Milky Way molecular-ring \ion{H}{2} regions.

\section{Observations}

Both G333.6$-$0.2 and W43 are extended compared to our 45--60$''$ KAO beam
and $80''$ LWS beam,
which means that integrating over a constant density model to calculate 
ICFs is not adequate.
Instead, we measured the fluxes at a number of positions at varying distances from the 
exciting stars to compare with the radial distributions 
of line flux computed from models to determine the source excitation and abundances.

\subsection{KAO Observations}

The observations were made with the Cryogenic Grating Spectrometer (CGS, Erickson et al. 1985)
on NASA's 0.91 m KAO telescope.
Data acquisition and reduction techniques are described by Colgan et al. (1993) 
and SCREH.
For the present observations, Saturn was used for calibrating 
the absolute fluxes and determining the relative detector responses.
The observed line and continuum fluxes for all positions are given in Table 1.

G333.6$-$0.2 was measured at its center and 
in three offset positions.
The data from the center position, C (RA $16^{\rm h} 22^{\rm m} 9\fs3$, Dec $-50\degr 6' 5''$,  both J2000),  
were taken with a 45$''$ beam
and were previously reported by Colgan et al. (1993). 
The three North positions (N1, N2, and N3 at 1$'$, 2$'$, and 3$'$ north of G333.6$-$0.2C)
were measured on 1993 May 11, 13, and 20, flying from Christchurch, New Zealand.
For these positions, a 60$''$ beam was used for all lines to provide good ratios 
to the [\ion{N}{2}] 205 \micron\ line, at whose wavelength the KAO has a diffraction 
size of $1.22 \lambda/D = 57''$.
The chopper angle was approximately east-west (position angle $\sim 90$\degr) 
and the throw was $\sim 8.5'$.
Because the [\ion{N}{2}] 122 \micron\ line is on the edge of a strong telluric H$_2$O line,
the observations of positions N1, N2, and the [\ion{N}{2}] 205 \micron\ line in Position N3
were taken at or near 43,000 feet altitude,
instead of the more usual 39,000--41,000 feet, where the other lines in Position N3 were measured.
Corrections for telluric H$_2$O absorption were made for both source and calibrator --- 
we estimate the zenith H$_2$O overburden 
to range from 4.4 to 5.2 \micron\ at 43,000 feet on May 11 and 13, and
3.9 \micron\ at 41,000 feet on May 20.

The three positions of W43 that had previously been studied by SCREH 
(W43N at RA $18^{\rm h} 47^{\rm m} 36\fs8$, Dec $-1\degr 55' 29''$; 
W43C at RA $18^{\rm h} 47^{\rm m} 36\fs0$, Dec $-1\degr 56' 39''$;
W43S at RA $18^{\rm h} 47^{\rm m} 38\fs5$, Dec $-1\degr 57' 34''$, all J2000)
in the [\ion{N}{3}], [\ion{O}{3}], and [\ion{S}{3}] lines were measured in the [\ion{N}{2}] lines 
on 1992 July 1, flying from NASA Ames. 
The 122 \micron\ line was measured in both $45''$ and $60''$ beams and 
the 205 \micron\ line was measured in a $60''$ beam.
The chopper angle was approximately east-west, the throw was $\sim 6'$, and
the zenith telluric H$_2$O overburden was $\sim 5$ \micron\ at 43,000 feet and
$\sim 2$ \micron\ at 45,000 feet.
We estimate that there are extra measurement uncertainties of order 20\% in addition to 
the statistical errors given in Table~1
for the W43 [\ion{N}{2}] 122 \micron\ lines.
These could result from clumpiness, pointing differences, or the effects 
of poor correction for deep atmospheric absorption.
The analyses of the various measurements produce different abundance measures,
as will be tabulated in the next sections.

\subsection{ISO Observations}

G333.6$-$0.2 was measured with the ISO Long Wavelength Spectrometer in LWS01 mode 
on 1997 March 20 (TDT No. 49001002, PI R. Rubin), exposure time 1054 s.
The line strengths for the [\ion{O}{3}] 52 and 88 \micron\ and [\ion{N}{3}] 57 \micron\ lines 
are given in Table 1 (the [\ion{O}{1}] 63 \micron\ line is also strong, but 
the [\ion{N}{2}] 122 \micron\ line was not detected owing to ISO's low spectral resolution
in LWS01 mode with the LW detectors).

\subsection{Extinction Corrections}

We corrected the measured line fluxes for extinction using the infrared
extinction function of Li \& Draine (2001). 
This function has a very similar wavelength dependence 
to the extinction function used by SCREH when 
normalized to Li \& Draine's extinction maximum at $\sim 9.5$ \micron.
Thus the FIR optical depths are characterized as being proportional to the 
maximum optical depth in the 9.5 to 10 \micron\ region, $\tau_{9.7}$.
Based on new information in the literature, we have updated these values from 
SCREH, with W43 having more extinction and G333.6$-$0.2 having less.

For three stars in the central cluster of W43, Blum et al. (1999) estimated that $A_K$ = 3.55, 3.52, and 3.63. 
Using the extinction law of Li \& Draine (2001),
we find that $A_K = 3.55$ corresponds to $\tau_{9.7} = 2.74$. 
Thus we increase the values of $\tau_{9.7}$ used by SCREH by 37\% to 
$\tau_{9.7} = 2.74$, 2.74, and 3.63 for W43N, W43C, and W43S, respectively.

For G333.6$-$0.2, Fujiyoshi et al. (2001) inferred an average value of $\tau_{9.7} = 1.5$ 
from their mid-IR spectropolarization measurements.
Consequently, we use $\tau_{9.7} = 1.5$ for the center position 
instead of SCREH's $\tau_{9.7} = 2.0$.
(This value corresponds to $A_{\rm H\alpha} = 14$ mag; 
a lower extinction yet for the center of G333.6$-$0.2 might be predicted, since the nebula 
was detected in H$\alpha$ by Churms et al. 1974.)
We use $\tau_{9.7} = 1.5$ for the North positions as well,
since there are no significant variations in the integrated \ion{H}{1} optical depths 
measured by Forster et al. (1987) at the C, N1, and N2 positions.

Following SCREH, an extinction uncertainty estimate of 20\% ($1 \sigma$) 
is propagated through all subsequent data analysis errors.

\section{Analytical Approach}

Because recombination, bremsstrahlung, and collisional excitation rates are
all proportional to the electron density squared, $N_e^2$, 
it has long been the practice to treat all abundance analyses 
in \ion{H}{2} regions and planetary nebulae (PNe) 
as though the surface brightness is, in fact, 
proportional to the emission measure, $EM = \int N_e^2 dl$,
where $l$ is the line-of-sight path through the nebula. 
This is valid so long as $N_e$ is substantially less than the critical density, $N_{crit}$, 
defined as the density at which the upper level of a transition is 
depopulated equally by radiative transitions and by collisions (e.g., Osterbrock 1989;
Rubin 1989).
Corrections for collisional de-excitation can be made if $N_e$ is known and approximately constant.
However, constant densities are unlikely to occur in real nebulae.
The effect of density variations is that the kernels in the integral 
over the nebular volume have different weighting functions for ions with
different amounts of collisional de-excitation.
Thus the flux ratios of such ions depend on these variable weighting functions
(see, e.g., Rubin 1989).
Here we consider the calculation of ratios of column densities of ions 
whose emission is proportional to $N_e$, as well as to $N_e^2$. 
The former is appropriate when $N_e \gg N_{crit}$.
These two methods give identical results for ion ratios when the density 
is constant over the whole observed volume. 
We will demonstrate the effects of density variations via a simple example 
and use the results to estimate how to handle the transition case where $N_e \sim N_{crit}$.

\subsection{Low Density \ion{H}{2} Regions}

For a given ion, a ``low density'' \ion{H}{2} region is one where $N_e \ll N_{crit}$
and a ``high density'' \ion{H}{2} region is one where $N_e \gg N_{crit}$;
thus any particular \ion{H}{2} region can have both low density and high density volumes,
depending on the value of $N_{crit}$ for the transition being studied.

In the optically thin case, the observed line flux is $F = \int j dl d\Omega$
where $j$ is the volume emissivity divided by $4\pi$ and 
$dl d\Omega$ is the volume element of the \ion{H}{2} region in the telescope beam, $\Omega$.
We define the normalized volume emissivity $\epsilon = 4\pi j/N_i N_e$ 
(Rubin et al. 1994; SCREH) 
where $N_i$ is the density of atoms in the particular ionization state
and the average $N_e$-weighted ionic density as
$<N_i> = \int N_i N_e dl d\Omega/\int N_e dl d\Omega$.
For the low density regime where $N_e \ll N_{crit}$, 
$\epsilon$ has very little dependence on $N_e$, and the line surface brightnesses (or fluxes) 
are proportional to $\int N_e^2 dl$ (or $\int N_e^2 dl d\Omega$).

Thus for this case we can write
\begin{equation}
F = \int {\epsilon\over 4 \pi} N_i N_e dl d\Omega \approx {\epsilon\over 4 \pi} <N_i> \int N_e dl d\Omega
\end{equation} 
and the ionic abundance ratio is
\begin{equation}
{<N_{i_1}> \over <N_{i_2}>} = {F_1/\epsilon_1 \over F_2/\epsilon_2}. 
\end{equation}
As $N_e$ approaches $N_{crit}$, collisional de-excitation of the line's upper level 
becomes more and more important, and the ratio of two lines from the same ionic species $N_i$
with different $N_{crit}$ can be used to estimate the electron density 
(e.g., Rubin 1989; Fig. 1a).

\subsection{High Density \ion{H}{2} Regions}

At high densities, the fractional population of the upper level approaches that of 
local thermodynamic equilibrium (LTE) and is a function of electron temperature, $T_e$, only.
In this regime, the line surface brightness is proportional 
to the column density, $CD = \int N_i dl$, the number of ions in the line of sight and
\begin{equation}
F = \int {(N_e \epsilon) \over 4 \pi} N_i dl d\Omega \approx {(N_e \epsilon) \over 4 \pi} \Omega \int N_i dl,
\end{equation}
where the quantity $N_e \epsilon$ is the emissivity per ion 
and has very little density dependence so long as there are no low density regions 
along the line of sight (no large density variations).
Then for identical beam areas $\Omega$, the column density ratio is
\begin{equation}
{CD_1 \over CD_2} = {F_1/(N_{e_1} \epsilon_1) \over F_2/(N_{e_2} \epsilon_2)}. 
\end{equation}
In fact, for abundance ratio calculations, one really wants to measure 
the actual number of ions in the line of sight, and {\it not} the number of ions weighted 
by some other, strongly varying parameter such as the electron density (as in equation (4)).
Thus it is desirable to estimate column densities for gaseous nebulae whenever feasible.

\subsection{Density Variations: An Illustrative Model}

The presence of density variations has a number of effects on the abundances that 
one infers from FIR line measurements.
We demonstrate these effects by the use of simple slab models, 
with an exponentially decreasing density $N_e = N_0 e^{-x}$
from a location, $x_0$, in the interior of the slab to an outer location, $x_{max}$. 
This density function was chosen because it gives similar weighting to both high and low 
density regimes (other types of models, e.g., Rubin 1989, could be used to 
illustrate the same effects). 
For simplicity, the example assumes no ionization stratification, $N_i = N_e$, and $T_e = 7000$ K; 
thus the only effects
on the calculated volume emissivities are those of the variation in density.
For all models, a very large value of $x_{max}$ was chosen such that $N_e = 0.01$ cm$^{-3}$.  
Each model had a different value of $x_0$ and $N_0$, 
ranging from $N_0 = 1$ cm$^{-3}$ to $N_0 = 10^6$ cm$^{-3}$.
The intensities, $I$, emitted in each line were calculated from
\begin{equation}
I =  \int _{x_0} ^{x_{max}} {\epsilon(x) \over 4 \pi} N_i N_e dx.
\end{equation}

Figure 1a shows how the electron density, $N_{est}$, 
can be estimated using calculated volume emissivity ratios from the observed 
[\ion{S}{3}] 19/33 \micron, [\ion{O}{3}] 52/88 \micron, and [\ion{N}{2}] 122/205 \micron\ line ratios,
with the [\ion{N}{2}] ratio being sensitive to low densities 
and the [\ion{S}{3}] ratio sensitive only to much higher densities.
Values of $N_{est}$ were computed from the ratios of the line intensities produced by the models; 
ratios of these estimated densities to the maximum density of each model, $N_0$,
are plotted in Figure~1b.
This figure shows that in the presence of density variations, even when the ions
coincide exactly in space, one does not infer the same density from the three ratios:
$N_{est}$(\ion{N}{2}) $ < N_{est}$(\ion{O}{3}) $< N_{est}$(\ion{S}{3}).

To understand the effects of density variations on estimated abundances, 
we calculate the estimated numbers of ions in the volume 
from the model line intensities and $N_{est}$
and compare these estimates with the true numbers of ions found by integrating 
$N_e$ or $N_e^2$ over the model line of sight.
These comparison ratios we call ``abundance measures''.
Numerically, we define the abundance measure for the low-density case where the 
line intensities are proportional to the emission measure as
\begin{equation}
AM_{EM} = {I \over \epsilon} \  \Bigg[ \int _{x_0} ^{x_{max}} N_i N_e dx \Bigg]^{-1} 
\end{equation}
and the abundance measure for the high-density case where the 
line intensities are proportional to the column density as
\begin{equation}
AM_{CD} = {I \over (N_{est}\epsilon)} \  \Bigg[ \int _{x_0} ^{x_{max}} N_i dx \Bigg]^{-1}.
\end{equation}

$AM_{EM}$ and $AM_{CD}$ are plotted in Figure 2 for each ion as functions of $N_{est}$.
Both abundance measures contain fewer ions than the input number,
but the regimes of reasonable validity for the two measures are different and depend on $N_{crit}$.
Fortunately, much of the discrepancy with the $N_e$ or $N_e^2$ volume integrals  
cancels when we take ratios
of the $EM$ or $CD$ abundance measures (equivalent to equations (2) and (4), respectively), 
as shown in Figure 3.
These plots suggest an approach for the intermediate cases where $N_{est} \sim N_{crit}$
and there are measurements of $N_e$ from (at least) two different density-sensitive line pairs:
 $N_{e_1}$ and $N_{e_2}$.

Using Fig. 3, we define three regimes as follows:
The low-density regime occurs when the estimated density is lower than the lowest
$N_{crit}$ for the lines involved (see Fig. 2 for $N_{crit}$) --- here we use 
the equations for emissivity $\propto N_e^2$ since the ionic ratios weighted by $N_e^2$
are close to the model input ratio (unity for the plots). 
The high-density regime occurs where the column density ratio is 
closer to the model input ratio 
than the ionic ratios weighted by $N_e^2$ --- this occurs when the estimated density 
is higher than the larger $N_{crit}$ of the more density-sensitive line pair.
At intermediate densities, both the low-density and high-density equations 
give abundances discrepant from the model input.
Considering this, in the intermediate regime we  
average the two cases and quote an uncertainty equal to the greater of 
one half the difference between the two abundance ratios and the statistical error.
For the low- and high-density regimes, we will use the equations for 
the ionic ratios weighted by $N_e^2$ or the column density ratios, respectively.
Unfortunately, it is extremely difficult to get a good estimate of the column density
of H$^+$ because the emissivities of H lines in \ion{H}{2} regions and PNe are always proportional to $N_e^2$ 
and the properly weighted density is impossible to determine.
Because of this, we are not able to measure abundance ratios with respect to hydrogen
for all the ions; instead 
we will obtain standard ``ionic ratios'' with respect to H$^+$ 
only for those ions whose lines have high $N_{crit}$
compared to the densities estimated from the observed line pairs. 
We will calculate both ionic ratios and column density ratios for the rest of the 
heavy element ions observed at FIR wavelengths 
and consider the density regime according to the line's $N_{crit}$ 
for estimating the final abundance ratio.

A significant difference between this method and previous abundance determinations 
from FIR lines is the increase in the size of the errors that occur when more 
than one density-sensitive line pair is measured in a high-density, compact \ion{H}{2} region.
{\it The smaller statistical errors that one obtains when only one density line pair is measured 
are an illusion.}
It is clear that measuring multiple density-sensitive line pairs is not just desirable
but is, in fact, required to obtain reliable abundance measurements from FIR lines 
(even though they may have large assigned errors).
We note that a similar situation occurs 
in optical abundance determinations from forbidden lines 
in that $T_e$ must be inferred from various line ratios
to obtain reliable abundances (e.g., Kennicutt, Bresolin, \& Garnett 2003).

\section{G333.6$-$0.2 and W43}

\subsection{Analytical Abundance Estimates}

The complete set of derived $N_e$ values and abundance ratios is given in Table 2.
The values of $T_e$ used were 6200~K (Caswell \& Haynes 1987) and 
6500~K (Wink, Wilson, \& Bieging 1982) for G333.6$-$0.2 and W43, respectively
(mid-and FIR lines are not sensitive to uncertainties in $T_e$).
The radio fluxes for G333.6$-$0.2 were estimated by integrating over the model
described in the next section at the appropriate distances from the center
(Figure 4) 
and the radio fluxes for W43 were taken from Lester et al. (1985) 
(the NRAO VLA Sky Survey, Condon et al. 1998, has lower fluxes, 
probably because the bright clumps in W43 are somewhat optically thick at 1.4 GHz,
and the ``snapshot'' observation of Balser et al. 1995 is missing extended flux).
The differences between the ionic abundance results presented here 
and in SCREH for the KAO data are due to 
different extinction corrections (\S 2.3),
different radio fluxes for G333.6$-$0.2,
and different collisional excitation cross sections for [\ion{S}{3}] (Tayal \& Gupta 1999).
The differences between the ISO and KAO abundances with respect to H$^+$ probably 
indicate the uncertainties in the radio flux estimates.
The N$^{++}$/O$^{++}$ ratios are the same to within the errors ($3 \sigma$), although 
some of the difference may be due to different beam sizes or pointing errors.

Estimated abundance ratios for N/S and S/H are given in Table 3, where we
assumed that the ionization fraction for sulfur $<$S$^{++}$/S$> = 0.9$ (a typical value
for the low-to-moderate excitation and high-metallicity models in Rubin 1985, 
SCREH, and the models computed in \S 5) 
and that the ionization fraction of $<$N$^{3+}$/N$>$ is so small
that N$^{3+}$ can be neglected.
The N/H ratio was derived by multiplying N/S times S/H.
Abundance estimates for the other elements are more uncertain 
because they require the use of models to estimate the necessary ICFs.

\subsection{Models of G333.6$-$0.2 and W43}

We computed \ion{H}{2} region models using 
the program NEBULA (Rubin 1968a; Rubin 1985; SCREH).
We have upgraded NEBULA since the discussion in SCREH in the following ways:

(1) The photoionization and recombination cross sections were extensively updated 
to include results from the Opacity Project (Seaton et al. 1992).
Fits to photoionization cross sections by Verner et al. (1996) were used for 
photoionization of C$^+$, C$^{++}$, N$^0$, N$^+$, N$^{++}$, O$^0$, O$^+$, Ne$^0$, Ne$^+$, 
S$^+$, S$^{++}$, S$^{3+}$, and Si$^+$.
The changes to the coefficients for ionization and recombination of iron 
are discussed by Rodr\'{\i}guez \& Rubin (2004).
(Changes to other elements were omitted because of the lack of corresponding Opacity Project 
recombination coefficients.)
Total recombination coefficients (including dielectronic recombination) were taken 
from the tabulations of Nahar \& Pradhan (1997) for C$^+$, C$^{++}$, and N$^{++}$;
Kisielius \& Storey (2002) for N$^+$;
Nahar (1999) for O$^0$ and O$^+$; 
Kisielius et al. (1998) for Ne$^+$;
Nahar (2000) for S$^{++}$; 
and Nahar (1995) for S$^+$ and Si$^+$. 
Charge exchange rate coefficients for reactions with H$^+$ were taken from the tabulation of
Kingdon \& Ferland (1996) with updates from the Oak Ridge National Laboratory web site 
``Charge Transfer Database''\footnote{\url{http://www-cfadc.phy.ornl.gov/astro/ps/data/}}
and coefficients for reactions with He$^+$ were taken from J. G. Wang et al. (2004, in preparation), 
with the data available from the same Oak Ridge web site.

(2) The collisional excitation cross sections now include 
N$^+$ and O$^{++}$ from Lennon \& Burke (1999),
N$^{++}$ from Blum \& Pradhan (1992),
Ne$^+$ from Saraph \& Tully (1994),
Ne$^{++}$ from McLaughlin \& Bell (2000),
S$^{++}$ from Tayal \& Gupta (1999),
S$^{3+}$ from Saraph \& Storey (1999) or Tayal (2000) for $T_e > 10,000$ K,
Ar$^+$ from Pelen \& Berrington (1995),
and Ar$^{++}$ from Galav\'{\i}s, Mendoza, \& Zeippen (1995, 1998).

(3) The diffuse radiation field calculation (Rubin 1968a) was modified to include
calculations at all radial grid positions instead of interpolation between selected positions.
Up to 20 iterations were used if needed to obtain uniform convergence 
(most positions converged within 10 iterations).

(4) Non-LTE stellar atmospheres with winds from Pauldrach, Hoffmann, \& Lennon (2001) were used.
Smith, Norris, \& Crowther (2002) have also computed model atmospheres using Pauldrach's code;
however, we did not use Smith et al.'s models because of the coarseness of their wavelength grid.

We used the set of model atmospheres, which were downloaded from Pauldrach's web 
site\footnote{\url{http://www.usm.uni-muenchen.de/people/adi/adi.html}},
consisting of models for Solar composition, dwarfs (D) and supergiants (S), and
\teff\ = 30, 35, 40, and 45 kK.
Although G333.6$-$0.2 and W43 have metallicities somewhat larger than that of the Orion Nebula,
the Orion Nebula abundances (e.g., Simpson et al. 1998) are sufficiently below 
the ``Solar'' composition used in the stellar atmosphere models  
that G333.6$-$0.2 and W43 
are probably closer in metallicity to the ``Solar'' composition models
than to anything greatly different.
Stellar atmospheres with higher metallicities have more winds, 
which produce more extreme ultraviolet (EUV) flux, but also more line
blanketing (less EUV); the difference is probably mainly for frequencies $> 4$ Ryd,
which are not important for \ion{H}{2} regions.
We computed models for W43 using both D and S models from Pauldrach et al. (2001) and also 
the non-LTE, plane-parallel model atmospheres without winds calculated by 
Lanz \& Hubeny (2003)\footnote{Available from \url{http://tlusty.gsfc.nasa.gov}}.
We used only the dwarf star model atmospheres for G333.6$-$0.2  
because it is clearly excited by a very young star cluster (Blum et al. 2001).

One result of the changes to the nebular atomic physics data 
is that the volume-averaged ionization fractions 
(SCREH; Rubin et al. 1994) for $<$O$^{++}$/O$>$ and $<$Ne$^{++}$/Ne$>$ 
predicted by the models are larger than previous predictions, 
whereas the fractions for $<$S$^{++}$/S$>$ and $<$N$^{++}$/N$>$ are approximately the same.
The differences are similar to typical uncertainties in ionization ratios
computed from observational data ($\lesssim 30$\%), 
which would make testing difficult unless the
data sample was large and of very high quality.
However, the choice of stellar atmosphere model makes a much larger difference
in the ionization fractions and ratios thereof; 
this choice has serious implications for estimates of \teff\ and ICFs.
We will compare various volume-averaged ionization fractions computed for different model
\ion{H}{2} regions as a function of stellar atmosphere model and 
discuss the need for winds further in \S 5.1.

\subsubsection{G333.6$-$0.2}

The initial model is that of a spherically symmetric, core-envelope \ion{H}{2} region
(Rubin, Hollenbach, \& Erickson 1993), with density 
decreasing from high values in the core to low values in the outer envelope, as
suggested by the estimated densities of Table 2.
Our initial density distribution for the G333.6$-$0.2 model computed here 
was derived by inverting the radio brightness 
temperature distribution, since the radio surface brightness is proportional to $\int N_e^2 dl$.
The radio brightness temperature distribution of G333.6$-$0.2, of necessity, has been estimated
by combining a variety of different measurements, none of which includes the whole
\ion{H}{2} region at a frequency high enough to insure that the emission is everywhere optically thin.
The highest spatial resolution radio maps in the literature consist of the following maps:
(1) a Parkes 64-m telescope map at 8.87 GHz with half-power beamwidth, HPBW = $2.5'$, 
made by McGee, Newton, \& Butler (1979);
because the map was made with a single dish, it covers the full spatial extent of G333.6$-$0.2
to a radius of $> 15'$;
(2) a Fleurs Synthesis Telescope map at 1.415 GHz with HPBW $\sim 50''$ by Retallack \& Goss (1980);
only the inner 5$'$ could be mapped, and the core is optically thick;
(3) an Australia Telescope Compact Array map at 3.4 cm (8.8 GHz) with resolution $1.6''$ by Fujiyoshi (1999);
only the inner $13''$ could be mapped, and the
peak brightness temperature of $\sim 4400$ K indicates that the central $2'' - 3''$ 
is not completely optically thin ($T_e \sim 6200$ K, Caswell \& Haynes 1987).
However, this last map is extremely important because it shows just how compact and bright
the central core of G333.6$-$0.2 really is. 
A compact bright core is also required by the NIR maps and the density-sensitive 
FIR line ratios.

Still lacking is a measurement of the transition region between the inner core and the envelope.
There is, however, a map of the warm dust emission made by the Midcourse Space Experiment 
(MSX; Price et al. 2001) at 21 \micron\ 
with $19''$ resolution.\footnote{Available at \url{http://irsa.ipac.caltech.edu}}
Although there is no certainty that warm dust has the same spatial structure as ionized gas 
(and it certainly does not have constant temperature), 
in the absence of an alternative 
we used it to interpolate between the core and envelope radio maps.

The total radio flux was taken from the measurements at the highest observed frequencies, $\nu$, 
because G333.6$-$0.2 shows evidence of being at least partially optically thick
at frequencies $< 10$ GHz.
We used the integrated flux of 93.8 Jy at 14.7 GHz measured by McGee \& Newton (1981).
The final estimated radio brightness distribution was renormalized to produce this
total integrated radio flux.
The estimated radio brightness temperature distribution, scaled $\propto \nu^{-2.1}$
as though it were optically thin at 5 GHz, 
and derived electron densities 
are shown in Figure 4.
The total radio flux requires a stellar ionizing photon luminosity 
of $10^{50}$ photons s$^{-1}$.
There may also be, of course, additional EUV photons that do not contribute to 
the \ion{H}{2} region photoionization because they are absorbed by dust.
Table 4 lists the relevant IR and radio observations, corrected for extinction,
including those mid-IR observations from the literature which include most of the core.

In addition to matching the radio brightness temperature distribution, 
the observations of G333.6$-$0.2 put a number of requirements on the parameters
of any model.
There must be a core of very high density to produce the bright radio
and mid-IR core and the high [\ion{O}{3}] 52/88 \micron\ and [\ion{S}{3}] 19/33 \micron\ line ratios 
seen in the center position.
The material at $2' - 4'$ must be of low density to produce the 52/88 \micron\ 
and 19/33 \micron\ line ratios of order unity ($N_e$ of order $150 - 400$ cm$^{-3}$).
The material further out must be of even lower density ($N_e \sim 40$ cm$^{-3}$) to make
a large contribution to the 122/205 \micron\ line ratio $\sim 2.5$.
The stellar \teff\ has to be high enough to produce O$^{++}$ and N$^{++}$ at all positions
(out to $> 3'$) but low enough so that there is ample N$^{+}$, Ar$^+$, and not 
much N$^{++}$, O$^{++}$, or S$^{3+}$ even in the center position. 
In addition, the He$^+$/H$^+$ radio recombination line (RRL) ratio 
is only 0.045 (McGee \& Newton 1981), 
indicative of a relatively low \teff\ ($\lesssim 35$ kK).

We computed over a hundred models of G333.6$-$0.2 with a variety of combinations of 
stellar atmosphere models and density filling factors, all models
reproducing the observed radio flux.
The abundance ratios used in the models were estimated from the fluxes in Table 4
(see also Table 3)
and range from 1.25 to 3.4 times the Orion Nebula abundances (Simpson et al. 1998).
Line fluxes were computed from the models 
by integrating over flat-topped circular beam profiles for the IR lines or
circular Gaussian beam profiles for the radio recombination lines
located at the projected distances of the observed positions from the \ion{H}{2} region center.

Our best example of a spherically-symmetric model is the G333.6$-$0.2 Core model in Table 4.
This particular model is shown because it predicts line ratios that
agree reasonably well with the density indicators ([\ion{S}{3}] 19/33 \micron, 
[\ion{O}{3}] 52/88 \micron, and [\ion{N}{2}] 122/205 \micron)
and the ionization indicator measured in the largest beam 
(He$^+$/H$^+$ RRLs measured by McGee \& Newton 1981).
The model uses a linear combination of 
0.9375 times the D-30 (30~kK Dwarf) plus 0.0625 times the D-35 stellar atmosphere models 
of Pauldrach et al. (2001) and has a volume filling factor of 0.5.
The contribution from the D-35 model, although seemingly small, 
is necessary to have enough high energy photons to ionize He and N 
to the observed ionization.
The given model combination only represents the shape of the ionizing spectrum 
and not the actual numbers or spectral types of the ionizing stars.

However, no model that we computed is a good fit --- the data seem to require 
stellar ionizing fluxes with both higher and lower average \teff.
The large fluxes of low-excitation ions, such as 
Ne$^{+}$ and Ar$^+$ in the center of G333.6$-$0.2 seem to require stars with low \teff\ 
($\lesssim 32$ kK was suggested by Rank et al. 1978).
However, the central density and optical depth are so high 
that when the exciting star is this cool, 
all the high energy photons that are capable of ionizing O$^+$ to O$^{++}$ 
are absorbed at small radii (0.87 pc, corresponding to $60''$)
and the models have essentially no O$^{++}$ at the radii corresponding 
to our three North positions.
Yet we observe O$^{++}$ at all four positions.
It would seem that an exciting star with a somewhat higher \teff\ ($> 35$ kK) is required
or a combination of stars including a much hotter star; 
however, for such models the $>35$ eV photons
required to ionize O$^+$ to O$^{++}$ still are confined in the central core
but the total [\ion{O}{3}] line fluxes are much larger than observed unless
the O/H, O/N, O/Ne, etc. ratios are unreasonably low.
On the other hand, in \ion{H}{2} regions ionized by stars with relatively low \teff, 
the N$^{++}$ Str\"omgren radius, $R_S$, is much larger than the O$^{++}$ $R_S$, 
leading to observed N$^{++}$/O$^{++}$ ratios that can be much larger than the actual N/O ratio
(e.g., Rubin et al. 1994; SCREH; Stasi\'nska \& Schaerer 1997).
The result of the differences in the N$^{++}$ and O$^{++}$ $R_S$ is
that the observed N$^{++}$/O$^{++}$ ratio should increase with projected distance from the center. 
This predicted increase is not observed in our N$^{++}$ and O$^{++}$ data
taken at the North positions.
Again, the observed behavior of this line ratio can be reproduced only by 
models with higher \teff\ stars.
In fact, all the models that we computed 
predict a much larger [\ion{N}{3}] 57/[\ion{O}{3}] 52 \micron\ ratio 
than is observed at any position. 
Agreement cannot be improved by increasing the O/H ratio 
(the models already predict more [\ion{O}{3}] 52 \micron\ flux than is observed)
or decreasing the N/H ratio (the model [\ion{N}{2}] lines fluxes agree with the observations).
A lower \teff\ seems to be required, but already there is too little O$^{++}$ 
in the North positions.
We conclude that a spherical model, even of varying density, cannot satisfy
the observations of both the very bright core and the doubly ionized lines
in the low density region at $2' - 4'$ from the center.

A possible solution might be some sort of non-spherical model with large clumps, or a blister model
(face-on from its symmetrical appearance, Hyland et al. 1980).
The NEBULA code for a blister model (Rubin et al. 1991) 
assumes that the density distribution is that
of the surface of a molecular cloud with some possible gradient of density
going into the cloud. The stellar photons ionize a hole on the edge of the cloud.
However, that axisymmetric model does not use a density function that allows for
a particularly high density immediately next to the stars, 
as is required for a model of G333.6$-$0.2.
Without a modification to the axisymmetric 2-D code, this code is not suitable.
A treatment that can handle a two- or three-dimensional density distribution appears 
to be required.

We next considered a quasi-blister model or two-component model, such as
was used by Simpson et al. (1986), Colgan et al. (1991), and Morisset et al. (2002)  
to model the Orion Nebula, K3-50, and G29.96$-$0.02, respectively. 
This would represent either a blister \ion{H}{2} region with one open side 
or an \ion{H}{2} region with very large clumps where ionizing photons escape 
through holes or paths between the clumps.
A linear combination
was made of two spherical models, one with the high density core and one
with an almost empty center to provide the doubly ionized ions seen at $2'-4'$.
The model with the empty center had a very large $R_S$ and therefore had to
be arbitrarily truncated in order that it be density bounded at about the
same radius as the $R_S$ for the Core model. 
An example of such a shell model, labeled ``Model for the North Positions'', 
is also given in Table 4; the model used the 
same stellar atmosphere as the Core model but the filling factor was unity.
Even here there are not enough photons to ionize O$^{++}$ at the N2 and N3 positions
and too much N$^{++}$ at the N1 position.
Possibly three or even four spherical models with different interior shell diameters 
and different scaling fractions for the volume of the model in the line of sight 
are needed to reproduce the observations at the four observed positions. 
Because of the arbitrariness of the procedure and the non-uniqueness of the result,
we did not calculate any of these multi-shell models.

The abundances have been only partially adjusted to produce the observed fluxes --- 
the immediate goal of the models
was to determine \teff, the filling factor, and other geometry factors.
Further adjustment to the abundances is not meaningful because the geometry still is 
not well determined, although the agreement of the He$^+$/H$^+$ RRL ratios 
with the observations indicates that \teff\ is not especially discrepant
(since most helium is primordial, the interstellar He/H ratio has not increased significantly 
as a result of Galactic chemical evolution). 
The uncertainty in the geometry factors 
makes it difficult to estimate a reliable abundance for oxygen.
However, if we consider just the Center and N1 positions,
we find that the observed fluxes would be matched by the model fluxes if
O/H$ \sim 3 \times 10^{-4}$, 
a value somewhat lower than that observed in the Orion Nebula.
This O/H abundance ratio is not reliable, but suggests 
that O/H is not especially high in G333.6$-$0.2.
As expected, since we observed the dominant ionization states of nitrogen and sulfur 
(N$^+$ and S$^{++}$) in this source, our final estimated abundances of these elements are 
those input to the models (Table 3).
On the other hand, it is difficult to think of a geometry 
that would produce the large [\ion{Ar}{2}] 6.98 \micron\ line flux
observed in the G333.6$-$0.2 core, given the number of high energy photons that are needed
to produce the observed [\ion{Ne}{3}] and [\ion{O}{3}] lines.
This discrepancy may indicate that the atomic data for argon need revision
(see also Morisset et al. 2004).

\subsubsection{W43}

Because the constant density model by SCREH for W43 is not a good 
match to the densities inferred from the observations (see Table 2), we calculated new models for it.
The appearance of W43 is that of a large ``T'' or incomplete ``D'' shape projected on the sky 
(Lester et al. 1985; Liszt 1995; 
Two Micron All Sky Survey image\footnote{2MASS, \url{http://irsa.ipac.caltech.edu}})
with a central star cluster near but not on the vertical element of the letter.
Because our three observed positions, including the one closest to the cluster,
all have similar densities and excitation (SCREH; Table 2), we interpret the geometry as
a cluster of ionizing stars surrounded by a cavity and then a shell of ionized gas, 
such that all three positions are at essentially the same physical distance from the stars.
Thus we model the  \ion{H}{2} region as 
a shell of ionized gas with interior radius = 1.2 pc ($0.75'$ at
a distance of 5.6 kpc (Reifenstein et al. 1970; Liszt 1995), scaled to 
a Galactic Center distance of 8 kpc).
The model $N_e$ drops smoothly from 770 cm$^{-3}$ at 1.2 pc to 420 cm$^{-3}$ 
at $R_S \sim 2.9$ pc, matching the average of the total observed flux of 77.5 Jy at 2 cm 
(Schraml \& Mezger 1969) and 97.4 Jy at 6 cm (Reifenstein et al. 1970) 
for an ionizing luminosity of $2.9 \times 10^{50}$ photons s$^{-1}$.
The densities and ionic abundance ratios of the three positions 
were averaged (weighted) for comparison with the models;
these comparisons are given in Table 5.

Three models were computed with the only difference being the input stellar atmosphere models.
The stellar atmosphere used for Model D is the sum of 0.8 times 
the D-35 model atmosphere from Pauldrach et al. (2001) plus 0.2 times 
the D-40 model atmosphere, 
the stellar atmosphere used for Model S is the sum 
of 0.6 times the S-35 model atmosphere plus 0.4 times the S-40 model atmosphere,
where D-35, D-40, S-35, and S-40 stand for Dwarf and Supergiant models with
$T_{\rm eff}$ equal to 35 kK and 40 kK, respectively,
and the stellar atmosphere used for Model T is the TLUSTY model atmosphere from 
Lanz \& Hubeny (2003) for $T_{\rm eff} = 35$~kK and log~$g = 3.50$.
The reason the S (supergiant) model atmospheres were also used is the presence
in W43 of a Wolf-Rayet star and two O giants/supergiants (Blum et al. 1999).
The combinations of stellar atmosphere models were adjusted until 
the \ion{H}{2} region models produced the observed average radio He$^+$/H$^+$ ratio 
(He/H = 0.10 for the nebular models) and the observed N$^{++}$/N$^{+}$ ratio.

To correct the observed ionic abundance ratios for unseen ionization states, 
we used the ionization fractions calculated from the model as shown in the Table~5 comments
and footnotes:
the averaged ionic abundance ratios were divided by the model ionic ratios, which are
inverse ICFs.
For example, the average observed O$^{++}$/S$^{++}$ ratio is 16.33 and Model S has 
$<$O$^{++}$/O$>$/$<$S$^{++}$/S$> = 0.391$.
Using this inverse ICF, we get O/S = 41.8 and O/H = O/S * S/H $ = 4.8 \times 10^{-4}$, using  
 S/H$ = 11.4 \times 10^{-6}$ for Model~S 
from the average S$^{++}$/H$^+ = 10.9 \times 10^{-6}$.
The main result to be noticed is that the N/H ratios are fairly similar for all models
at 1.2 to $1.5 \times 10^{-4}$, even though the O/H ratio varies by a factor of 2.5.  
The N/S ratio calculated by averaging N/S from $N^+$/S$^{++}$ and N$^{++}$/S$^{++}$ is
9.9 to 10.4, very similar to the result from summing (N$^{++}$ + N$^+$)/S$^{++}$ 
found in Table~3.

The O/H variation is caused by the model $<$O$^{++}$/O$>$ ionization fractions,
as seen in Table 5 in  
the $<$O$^{++}$/O$>$/$<$S$^{++}$/S$>$ and $<$N$^{++}$/N$>$/$<$O$^{++}$/O$>$ ratios. 
No doubt the reason is that the Supergiant model atmospheres have less flux from 35 to 41 eV
than the Dwarf model atmospheres.
Pauldrach et al.'s (2001) Supergiant models have much higher mass loss rates than 
their Dwarf models, so the difference must be caused by more line blanketing 
at these energies by the gas in the wind.
On the other hand, the static, plane-parallel, non-LTE atmospheres of Lanz \& Hubeny (2003),
which have no winds, have much lower fluxes from $\sim 40$ to $\sim 50$ eV.

All three W43 models adequately reproduce the \teff-indicating indices: 
the He$^+$/H$^+$ and N$^{++}$/N$^+$ ratios.
The range of implied abundances in Table 5, though, indicates the large uncertainties 
resulting from determination of abundances through model-produced ICFs,
particularly when different stellar atmosphere models are used to compute the ICFs.
We prefer the values of N/H derived from comparisons with sulfur, rather than 
those derived from oxygen.
The estimated N/O ratio is 0.32 to 0.42 from ICFs computed with models D and S  
(not very different from the 0.34 estimated by SCREH),
but is as low as 0.16, similar to that of the Orion Nebula (Rubin et al. 1991),
from the ICFs computed with model T.
The reason for the wide range is not the abundances of nitrogen or sulfur, 
which are reliable because they do not require large ICFs.
The problem is the total abundance of oxygen:
the model ICFs produce O/H ranging from 2.8 to 8.3 $ \times 10^{-4}$,
whereas the predicted O/H ratio is $7.2 \times 10^{-4}$
for a Galactic abundance gradient of $d($log O/H$) /dR = -0.06$ dex kpc$^{-1}$ 
(Henry \& Worthy 1999), given the Orion Nebula's location at $R_G \sim 8.4$ kpc
and O/H = 4.0 to $4.4 \times 10^{-4}$ (Rubin et al. 1991; Esteban et al. 1998).
In fact, one expects that the oxygen abundance for W43 should be much higher than that of
the Orion Nebula,  
given W43's low $T_e = 5410$ to 6500~K 
(Subrahmanyan \& Goss 1996; Wink, Wilson, \& Bieging 1983; and references in both papers).
For this source, as for G333.6$-$0.2, 
we conclude that we still do not have a good estimate for the abundance of oxygen,
the most abundant heavy element and the most important contributor to cooling in \ion{H}{2} regions.

\section{Discussion and Conclusions}

\subsection{Stellar Atmosphere Models}

As was discussed above, many of the ionization correction factors are critically dependent
on the choice of stellar atmosphere model.
This can be demonstrated either by plotting ratios of the numbers of photons
able to ionize various ions as a function of \teff\ or
by plotting ionization fractions or ICFs calculated from models;
here we do the latter because ICFs are more readily applicable to abundance computation.
This is illustrated in Figure 5, where we plot 
the ICF ratios $<$N$^{++}$/N$>$/$<$O$^{++}$/O$>$, $<$O$^{++}$/O$>$/$<$S$^{++}$/S$>$, and
$<$O$^{++}$/O$>$ versus the $<$N$^{++}$/N$>$/$<$N$^{+}$/N$>$ ratio,
which indicates excitation without the necessity for any abundance assumption.
The models from Rubin (1985) that are plotted in Fig. 5 
all have total nucleon density $= 10^3$ cm$^{-3}$, 
ionizing luminosities of $10^{49}$ to $10^{50}$ photons s$^{-1}$,
and Orion Nebula metallicities;
the models from SCREH had similar densities and ionizing luminosities but
metallicities ranging from 1 to 2 times Orion nebula metallicities.
The models from both Rubin (1985) and SCREH were computed with the versions of NEBULA 
available at the time
and both used LTE stellar atmosphere models from Kurucz (1979). 
In addition to the models described in \S 4, constant nucleon density ($ = 10^3$ cm$^{-3}$)
spherical models with 
ionizing luminosities of $10^{49}$ photons s$^{-1}$ were computed and plotted.
These recent NEBULA models used stellar atmosphere models from either
Pauldrach et al. (2001), Lanz \& Hubeny (2003),  
or contemporary\footnote{\url{http://kurucz.harvard.edu/}}
LTE stellar atmosphere models described by Kurucz (1992).
The stellar atmosphere models had ``Solar'' abundances and the nebular models
had 1.5 times Orion Nebula abundances (Simpson et al. 1998) such that the model abundance ratios 
(all times $10^{-4}$) are
C/H = 3.75, N/H = 1.02, O/H = 6.0, Ne/H = 1.5, S/H = 0.11, and Ar/H = 0.04.
Because they are widely used, we also plot ICFs from the
models of Stasi\'nska \& Schaerer (1997).
For these models both the stellar and nebular abundances were ``Solar''. 
We note that some of the elemental abundances considered to be ``Solar'' are changing with time; 
for example the ``Solar'' O/H ratio used by Stasi\'nska \& Schaerer (1997) 
was O/H$ = 8.51 \times 10^{-4}$ (Anders \& Grevesse 1989), whereas 
currently, the ``Solar'' O/H ratio is thought to be $5.45 \times 10^{-4}$ (Holweger 2001).
Mart\'{\i}n-Hern\'andez et al. (2002b), Smith et al. (2002), and Morisset (2004) 
discuss the effects of stellar abundances on \ion{H}{2} ionization.
Details of the physics used, though, are much more important than the exact ``Solar'' abundances,
as we will see.

Unless an \ion{H}{2} region can be observed at both optical and infrared wavelengths, 
observers cannot measure the abundances of N, O, Ne, S, and Ar without making corrections
for unseen ionization states for some elements.
For optically obscured \ion{H}{2} regions like W43 and G333.6$-$0.2, the most significant 
missing ion is O$^+$, since oxygen is the most abundant heavy element and since 
these low excitation \ion{H}{2} regions are believed to have most of their oxygen as O$^+$.
Rubin et al. (1994) and SCREH assumed that they knew the O/S ratio 
and used the observed O$^{++}$/S$^{++}$ ratio to estimate the ICF for their
measured N$^{++}$/O$^{++}$ ratios, since they did not have any other line pair
that could be used to indicate excitation.
In this paper we use our observed N$^{++}$/N$^{+}$ ratios to estimate \teff; 
however, we see from Fig. 5a how dependent the N$^{++}$/O$^{++}$ 
ICF is on the stellar atmosphere model.

Mart\'{\i}n-Hern\'{a}ndez et al. (2002a) took advantage of their ISO SWS measurements 
of the Ne$^{++}$/Ne$^{+}$ ratios to estimate ICFs without the need for 
assuming {\it a priori} 
some abundance ratio; unfortunately, the large beam size difference
between the SWS and LWS instruments means that their observed Ne$^{++}$/Ne$^{+}$ ratios
may not always be appropriate for estimating the excitation of O$^{++}$ and N$^{++}$.
Moreover, Ne$^{++}$ requires ionizing photons $> 41$ eV, and it is at these high energies 
that the stellar atmosphere models are especially divergent, as we now discuss.

An important test of the validity of stellar atmosphere models of hot stars is
whether \ion{H}{2} region models produced with these atmospheres 
predict line fluxes that agree with observations.
Except for a few rare instances, 
this is the only way that the EUV fluxes of the models can be 
tested, since interstellar absorption by hydrogen predominantly prevents the EUV fluxes from 
hot O stars from reaching the Earth.
Here we discuss the stellar fluxes that can doubly ionize neon (photon energy $ > 41$ eV).
This has been called the [\ion{Ne}{3}] problem (e.g., Sellmaier et al. 1996) because 
it has been observed that the Ne$^{++}$/O$^{++}$ ratio is relatively constant
over a large range of \ion{H}{2} region excitation 
(SCREH; Stasi\'nska \& Schaerer 1997 and references therein; Kennicutt et al. 2003), 
contrary to predictions of models that 
used LTE stellar atmospheres (e.g., SCREH).
Proposed solutions have involved using non-LTE stellar atmospheres, usually with winds 
(Rubin, Kunze, \& Yamamoto 1995; Sellmaier et al. 1996; Stasi\'nska \& Schaerer 1997).
Figure 6 shows the observations of SCREH along with the predictions of the models used in
Fig. 5 and the G333.6$-$0.2 and W43 models described in \S4.
Models calculated with black bodies for the stellar atmosphere spectrum are also plotted
since Morisset et al. (2004) found that such models give a reasonable fit to the mid-IR ISO observations.
Like the new nebular models plotted in Fig. 5, these models also have spherical geometry, 
constant nucleon density of $10^3$~cm$^{-3}$, $10^{49}$ ionizing photons, 
and 1.5 times the Orion Nebula abundances  listed in Table 3.

Surprisingly, in Fig. 6 only the models of Stasi\'nska \& Schaerer (1997) 
reproduce the observations over the total range of observed O$^{++}$/S$^{++}$, 
although if Ne/O is as large as 0.25 in all \ion{H}{2} regions as it is in the Orion Nebula
(Simpson et al. 1998; Table~3), 
these \ion{H}{2} region models produce too high a Ne$^{++}$/O$^{++}$ ratio
because of the very large EUV fluxes in Schaerer \& de Koter's (1997) stellar atmosphere models.
Moreover, Schaerer \& de Koter's models have too much EUV flux 
for reproducing other observations (e.g., Smith et al. 2002).
Earlier, Sellmaier et al. (1996) had demonstrated that the [\ion{Ne}{3}] problem was 
supposedly solved when they obtained a good fit to the data 
by using non-LTE atmospheres with winds computed with Pauldrach's code as it then existed.
However, our \ion{H}{2} region models with non-LTE stellar atmospheres with 
winds from Pauldrach et al. (2001) taken from both A. Pauldrach's web site and 
from the stellar atmosphere models produced by Smith et al. (2002) 
using Pauldrach's WM-basic code predict much lower Ne$^{++}$/O$^{++}$ than observed
for Dwarf atmospheres with \teff\ $< 40$ kK and for Supergiant atmospheres 
with \teff\ $< 35$ kK. 
Our \ion{H}{2} region model produced with a 35~kK supergiant atmosphere from F. Sellmaier 
(private communication in 1995) 
lies above the Pauldrach et al. Supergiant line but not as high as the model with 
the 35~kK supergiant atmosphere in Sellmaier et al.'s (1996) Figure 2,
even though we are now assuming Ne/O = 0.25 (Simpson et al. 1998) instead of 0.2025.
Part of the difference of the current nebular model predictions 
with previous models is the higher O$^{++}$ ionization relative to Ne$^{++}$ and S$^{++}$ 
obtained with the Opacity Project cross sections described in Section 4.
The model computed with Sellmaier's (1995) atmosphere has higher ionization than 
the model computed with Pauldrach et al.'s (2001) S-35  atmosphere because 
it used Orion Nebula abundances (as in Sellmaier et al. 1996) instead of
the higher metallicity (1.5 times the Orion Nebula abundances of Table 3)
used for the rest of the models plotted here.
The result, though, is a return of the mismatch between the stellar atmosphere models 
and the EUV fluxes of real stars, with possible additional discrepancies for 
the ionization, recombination, and collisional excitation cross sections 
for oxygen, neon, and sulfur.
Moreover, we see that all the models computed without winds 
do a poorer job of reproducing the observations than both the models computed with 
non-LTE atmospheres with winds and 
the models computed with simple black bodies for stellar atmospheres
(for this reason we give low weight to the T model for W43 in Table 5).
This probably means that the codes producing the stellar atmosphere models 
used for computation of \ion{H}{2} region models 
should include the effects of stellar winds.

\subsection{Abundance Ratios}

At the low excitations seen in inner Galaxy \ion{H}{2} regions, 
the predicted flux of the O$^{++}$ lines is extremely
sensitive to the stellar atmosphere, but if the models are accurate, the O/H ratio 
of both G333.6$-$0.2 and W43 
could be as low as $3 - 5 \times 10^{-4}$, substantially lower than would be predicted 
from the observed Orion Nebula abundance and an O/H abundance gradient 
of $-0.06$ dex kpc$^{-1}$ (Henry \& Worthey 1999).
This low O/H ratio is hard to understand because of the low $T_e$ measured in inner Galaxy 
\ion{H}{2} regions from radio recombination lines.
There would need to be some source of systematic error for 
the O/H ratio to be higher than these estimates: i.e., what would be required is that   
the excitation of oxygen in the two \ion{H}{2} regions is lower than 
estimated in this paper from He$^+$/H$^+$ recombination lines and N$^+$ and N$^{++}$ forbidden lines.
Possible reasons that the G333.6$-$0.2 and W43 excitation might be overestimated 
are as follows:
(1) The abundance of N$^+$ is underestimated because the [\ion{N}{2}] line fluxes are undermeasured 
owing to telluric absorption at 122 \micron\ or diffraction at 205 \micron.
This is a particular problem for W43, where the two measurements of 
the [\ion{N}{2}] 122 \micron\ line in different apertures 
can be used to produce very different electron densities 
and N$^{++}$/N$^{+}$ abundance ratios (0.6 to 3.2 in Table 5).
(2) The estimated N$^{++}$/N$^{+}$ ratio is too high because some of the estimated N$^+$ is missing 
owing to inadequate correction for density variations.
(3) The estimated N$^{++}$/N$^{+}$ ratio is too high because of 
incorrect collisional excitation cross sections for the pertinent energy levels.
(4) The models predict too much O$^{++}$/O versus N$^{++}$/N$^{+}$. 
The spread of O$^{++}$/O ratios in Fig. 5c shows that any excessive O$^{++}$/O 
is due to model atmospheres
and not due to the $\sim 30$\% higher oxygen ionization resulting from the Opacity Project cross sections.
(5) The ionization of helium might be overestimated 
if the He/H abundance ratio is actually greater than the assumed 0.10.
We conclude that determinations of the O/H abundance ratio in low excitation 
\ion{H}{2} regions from FIR data have large uncertainties at this time, 
with the largest contributor to the uncertainty being the choice of stellar atmosphere models, 
with some possible contribution from the atomic data used in the nebular models.

We have demonstrated that the abundances of nitrogen and sulfur can be obtained from FIR observations
and the ratio of N/S can be obtained when both ionization states of N$^{+}$ and N$^{++}$ are measured.
(Simpson et al. 1998, Mart\'{\i}n-Hern\'{a}ndez et al. 2002a and 2003, and Giveon et al. 2002 
also show that the abundances of Ne/H and Ar/H, as well as S/H, 
can be determined from mid-IR observations 
with little bias owing to excitation or abundance.)
Compared to SCREH, the S/H abundance ratio is higher in G333.6$-$0.2, owing to the
new data at the additional North positions, but lower in W43
because of the higher collisional excitation cross sections for sulfur.
The N/H ratio is about 50\% larger in G333.6$-$0.2, owing to the new data at the North positions,  
but a factor of 2 smaller in W43, owing to the smaller ICF for N$^+$.
However, since the estimated O/H ratio is also much smaller in W43, 
now that we are estimating the ionization from N$^{++}$/N$^+$ 
and not from O$^{++}$/S$^{++}$ with the assumption of a constant O/S ratio as SCREH did, 
the estimated N/O ratios agree with SCREH for both G333.6$-$0.2 and W43.
With this new analysis, 
the S/H and N/O ratios are also consistent with SCREH's measurements.

\subsection{Conclusions Regarding Systematic Errors}

There are several possible sources of systematic error 
that must be considered when measuring abundances in obscured, low-excitation \ion{H}{2} regions,
that is, all the \ion{H}{2} regions in the inner Galaxy.
The impact of these systematic errors is that there is a very large uncertainty 
in the total metallicity and important abundance ratios like N/O.
These sources of systematic error are the following:

(1) The most abundant heavy element ion
in low-excitation \ion{H}{2} regions, O$^+$, has no bright infrared lines.
The consequence is that the abundance of this critically important element 
must be estimated from the abundance of an ionization state 
with fractional abundance $< 0.5$.

(2) Nitrogen, the secondary element which one needs for studies of galaxy chemical evolution,
has two important ionization states: N$^+$ as well as N$^{++}$.
Fortunately, both species have measurable FIR lines, 
but {\it both} must be measured to determine the excitation.

(3) Moreover, both FIR N$^+$ lines have greatly different values of $N_{crit}$ from 
those of S$^{++}$ or N$^{++}$.
Thus one needs to be aware of and compensate for the ever-present density variations.

(4) The average density could be quite different in the various ionization zones,
as shown by the different electron densities derived from $N_e$-diagnostic line pairs: 
low ionization (N$^+$), 
intermediate ionization (S$^{++}$), and high ionization (O$^{++}$).

(5) ICFs estimated from models have major uncertainties owing to uncertainties 
in the stellar atmosphere models and 
possibly the atomic data used in the \ion{H}{2} region models.

It is clear that the abundances of primary (oxygen, neon, or sulfur) 
and secondary (nitrogen) elements will need to be measured in many more
inner Galaxy \ion{H}{2} regions before we understand the chemical evolution 
of the Milky Way.
In the near future, measurements of N$^{++}$, N$^+$, O$^{++}$, Ne$^{++}$, and S$^{++}$ 
will be possible from  
the Stratospheric Observatory for Infrared Astronomy, SOFIA.
We recommend that future observers measure excitation-sensitive lines,
such as Ne$^+$ and Ne$^{++}$, as well as both N$^+$ and N$^{++}$ 
to determine the \ion{H}{2} region ionization state
needed to estimate the ICF for O$^+$.
It is important that a sufficient number of lines be measured in each \ion{H}{2} region 
(and extended \ion{H}{2} regions mapped) so that detailed models of the \ion{H}{2} region
can be made. Discrepancies between the predicted line fluxes from the models 
and the observations can then be used to indicate EUV energy regions where 
the stellar atmosphere models may need revision.

\acknowledgements

We thank the staff of the KAO for their support while these observations were carried out
and J. Baltz, S. Lord, and A. Rudolph for their assistance with 
the observations and scheduling.
We especially thank those researchers who make their stellar atmosphere models and other data 
available to the public via the World Wide Web.
We thank D. Hollenbach and F. Witteborn for reading the manuscript
and the referee for his comments, which improved the paper.
This research made use of data products from the Midcourse Space 
Experiment (MSX).  Processing of the data was funded by the Ballistic 
Missile Defense Organization with additional support from the NASA 
Office of Space Science.  This research has also made use of the 
NASA/IPAC Infrared Science Archive, which is operated by the 
Jet Propulsion Laboratory, California Institute of Technology, 
under contract with the National Aeronautics and Space Administration.
The models were run on Cray computers at NASA/Ames (Consolidated Supercomputer Management Office) 
and at JPL; the latter was provided by funding 
from JPL Institutional Computing and Information Services and
the NASA Offices of Earth Science, Aeronautics, and Space Science.
J. P. S. acknowledges support from NASA Ames Research Center Co-operative Agreements NCC2-1134
and NCC2-1367.  
R. H. R. acknowledges support from a NASA Long-Term Space Astrophysics grant,
which funds his Co-operative Agreement NCC2-9021.

\clearpage

\begin{figure} 
\plotone{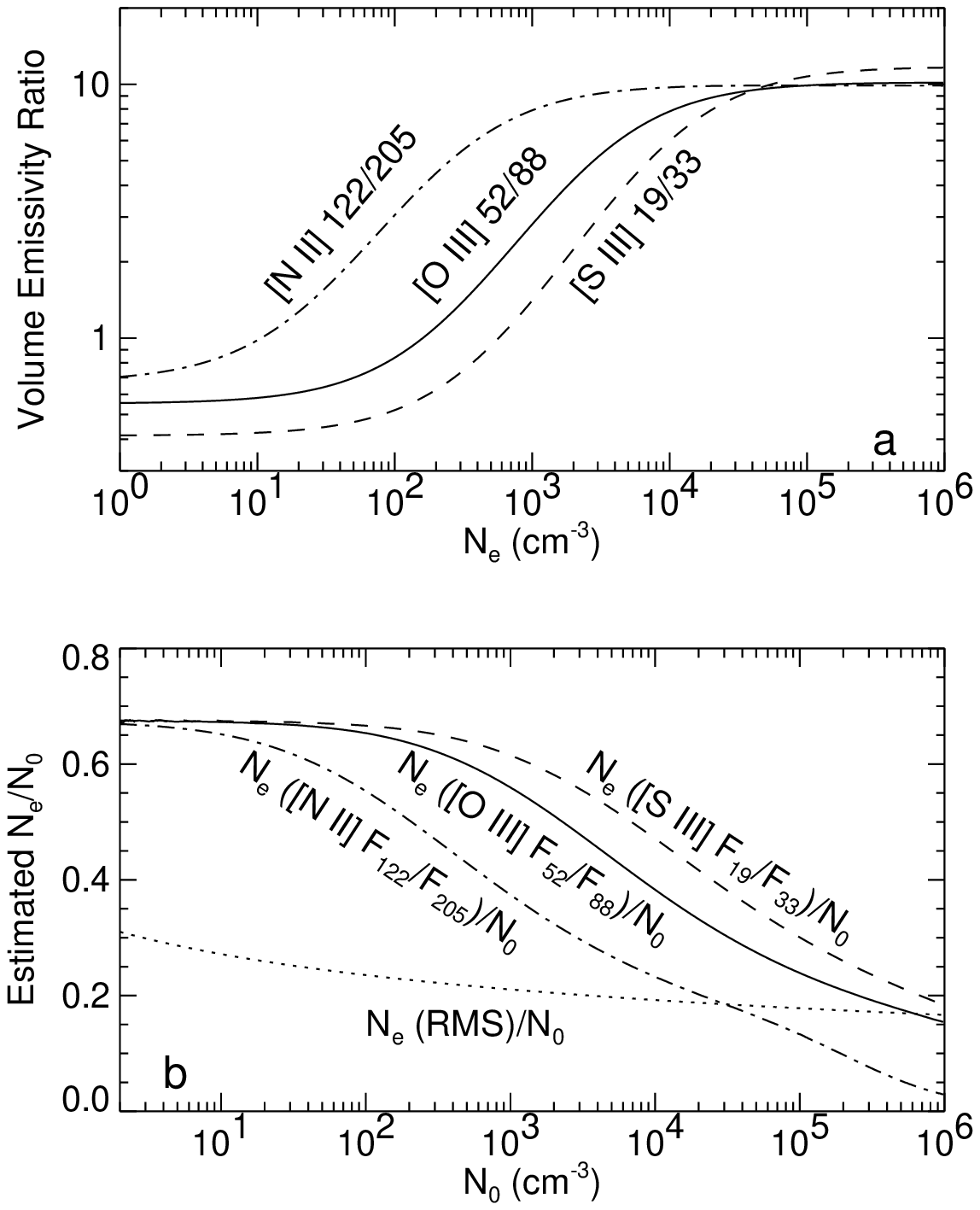}
\caption{($a$) Ratios of volume emissivities for line pairs of the 
electron-density-sensitive lines [\ion{S}{3}] 19/33 \micron, [\ion{O}{3}] 52/88 \micron,
and [\ion{N}{2}] 122/205 \micron\ plotted as a function of electron density, $N_e$, 
for electron temperature, $T_e$, of 7000 K.
($b$) Line intensities were calculated for slab models with $N_e = N_0 e^{-x}$ by 
integrating the volume emissivity of each line from $x_0$ to $x_{max}$ (see text). 
Values of $N_e$ were estimated from the abscissa of Fig.~1a 
by assuming that the ratios of the line intensities could be used for the ordinate. 
These estimated values of $N_e$ divided by $N_0$ are plotted versus $N_0$. 
The root-mean-squared (RMS) density divided by $N_0$ is also plotted (dotted line).}
\end{figure}
 
\clearpage
 
\begin{figure}
\plotone{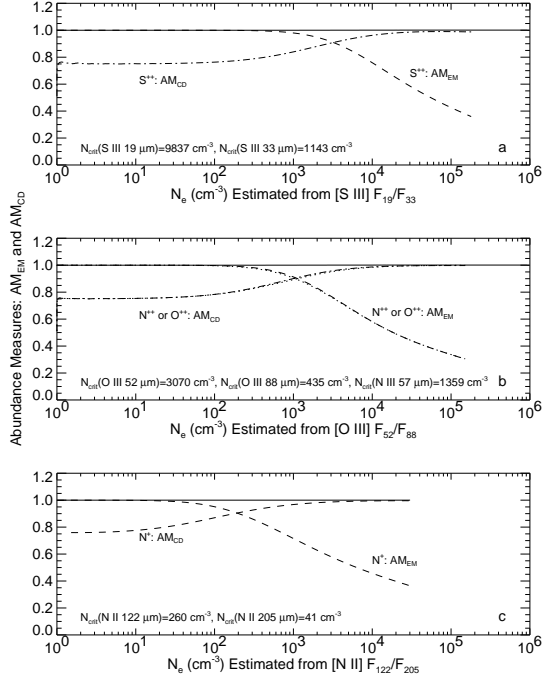}
\caption{
Normalized abundance measures calculated for slab models with exponentially decreasing density 
(see text) plotted as a function of the densities 
estimated from the ratios of the density-sensitive lines (see Fig.~1a).
The abundance measures, $AM_{EM}$ and $AM_{CD}$, are defined as the calculated line intensity 
divided by the normalized volume emissivity or the emissivity per ion, respectively,
and are normalized by dividing by the emission measure or
by the column density of ions, respectively (see text).
Because preferred abundance measures for each ion depend on the density, 
critical densities, $N_{crit}$, are also tabulated for $T_e = 7000$ K.
($a$) S$^{++}$ measured from [\ion{S}{3}] 19 and 33 \micron\ lines. 
($b$) O$^{++}$ and N$^{++}$ measured from [\ion{O}{3}] 52, [\ion{O}{3}] 88, and 
[\ion{N}{3}] 57 \micron\ lines. Note the near  coincidence of the N$^{++}$ and O$^{++}$ abundance measures.
($c$) N$^{+}$ measured from [\ion{N}{2}] 122 and 205 \micron\ lines.}
\end{figure}
 
\clearpage
 
\begin{figure}
\plotone{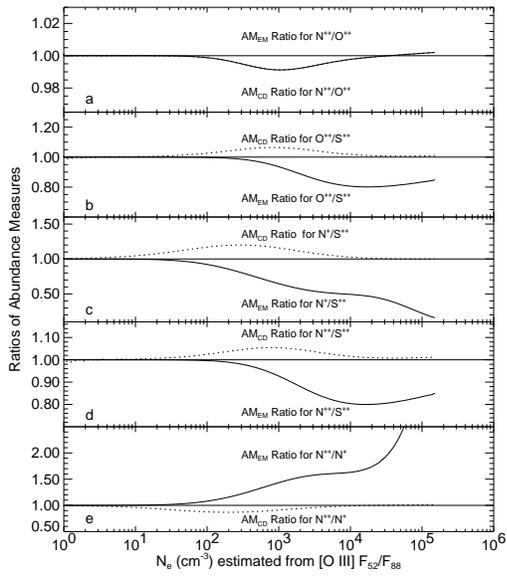}
\caption{Ratios of the abundance measures for the ions shown in Fig. 2 
plotted as a function of estimated $N_e$
for lines of sight with varying amounts of high-density gas.
The heavy line is $AM_{EM}$ and the dotted line is $AM_{CD}$
(the light line is unity, for reference).
($a$) N$^{++}$/O$^{++}$. The lines for $AM_{EM}$ and $AM_{CD}$ coincide.
($b$) O$^{++}$/S$^{++}$.
($c$) N$^{+}$/S$^{++}$.
($d$) N$^{++}$/S$^{++}$.
($e$) N$^{++}$/N$^{+}$.}
\end{figure}
 
\clearpage
 
\begin{figure}
\plotone{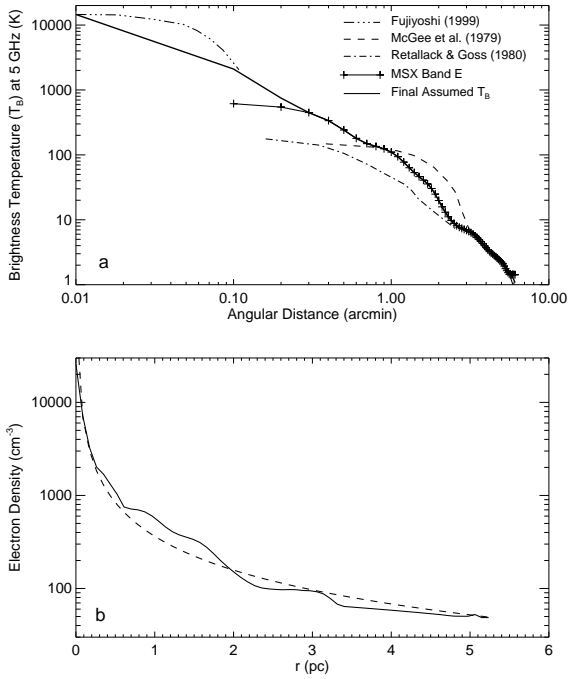}
\caption{($a$) Azimuthally-averaged radio brightness temperatures for G333.6$-$0.2 plotted
as a function of angular distance on the sky from the central peak.
The radio brightness temperatures were all scaled to the same frequency, 5 GHz, using the 
formulation of Mezger \& Henderson (1967) and ignoring optical depth effects.
($b$) $N_e$, derived by inverting 
the radio brightness temperature distribution shown in Fig. 4a, 
is plotted as a function of distance, $r$, from the central star cluster.
A spherical \ion{H}{2} region at a distance of 3~kpc was assumed.
The overall form of this density function is $N_e \propto r^{-1.2}$,
as shown by the dashed line, which was fit by least squares.
}
\end{figure}
 
\clearpage
 
\begin{figure}
\plotone{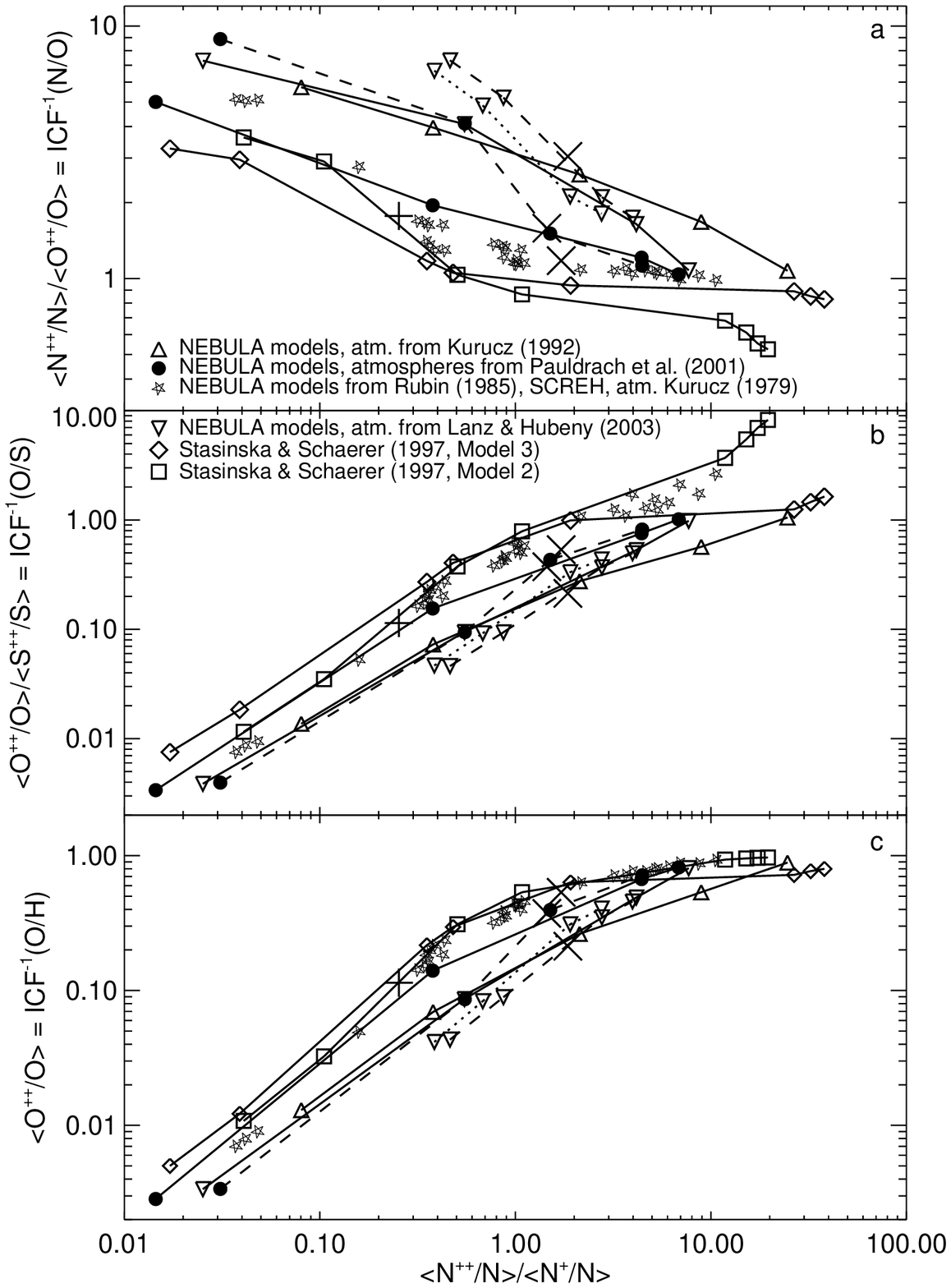}
\end{figure}
\notetoeditor{This figcaption goes with Figure 5;  however, the figure would have to be shrunk by 
a factor of 2 to fit on the same page with the caption. This makes it too small to be read easily.
Consequently, I printed them on 2 different pages. LaTeX puts the caption ahead of the figure.}
\figcaption{Ratios of ionization fractions plotted for spherical \ion{H}{2} region models.
The models with LTE stellar atmosphere models are those from Rubin (1985) and SCREH, 
who used model atmospheres from Kurucz (1979), 
and the new NEBULA models with contemporary atmosphere models described by Kurucz (1992)
(\teff\ = 33, 35, 37, 40, and 45 kK).
The \ion{H}{2} region models with non-LTE stellar atmosphere models 
are those from Stasi\'nska \& Schaerer (1997), 
who used atmosphere models from Schaerer \& de Koter (1997), and
the NEBULA models calculated for this paper, where we used non-LTE atmosphere models
from Pauldrach et al. (2001) and Lanz \& Hubeny (2003) 
(\teff\ = 30, 35, 40, and 45 kK).
For the \ion{H}{2} region models calculated with Pauldrach et al. atmospheres, 
the solid line connects models with Dwarf atmospheres 
and the dashed line connects models with Supergiant atmospheres.
For the \ion{H}{2} region models calculated with Lanz \& Hubeny atmospheres, 
the solid line connects models with atmospheres with log $g = 4.0$
and the dotted and dashed lines connect models with atmospheres with log $g = 3.0$ to 3.5 and 
with Lyman continuum luminosities of $10^{49}$ and $10^{50}$ photons s$^{-1}$, respectively.
The notation in the legend applies to all parts of the figure.
($a$) $<$N$^{++}$/N$>$/$<$O$^{++}$/O$>$ versus $<$N$^{++}$/N$>$/$<$N$^{+}$/N$>$.
($b$) $<$O$^{++}$/O$>$/$<$S$^{++}$/S$>$ versus $<$N$^{++}$/N$>$/$<$N$^{+}$/N$>$.
($c$) $<$O$^{++}$/O$>$ versus $<$N$^{++}$/N$>$/$<$N$^{+}$/N$>$.
}
 
\clearpage
 
\begin{figure}
\plotone{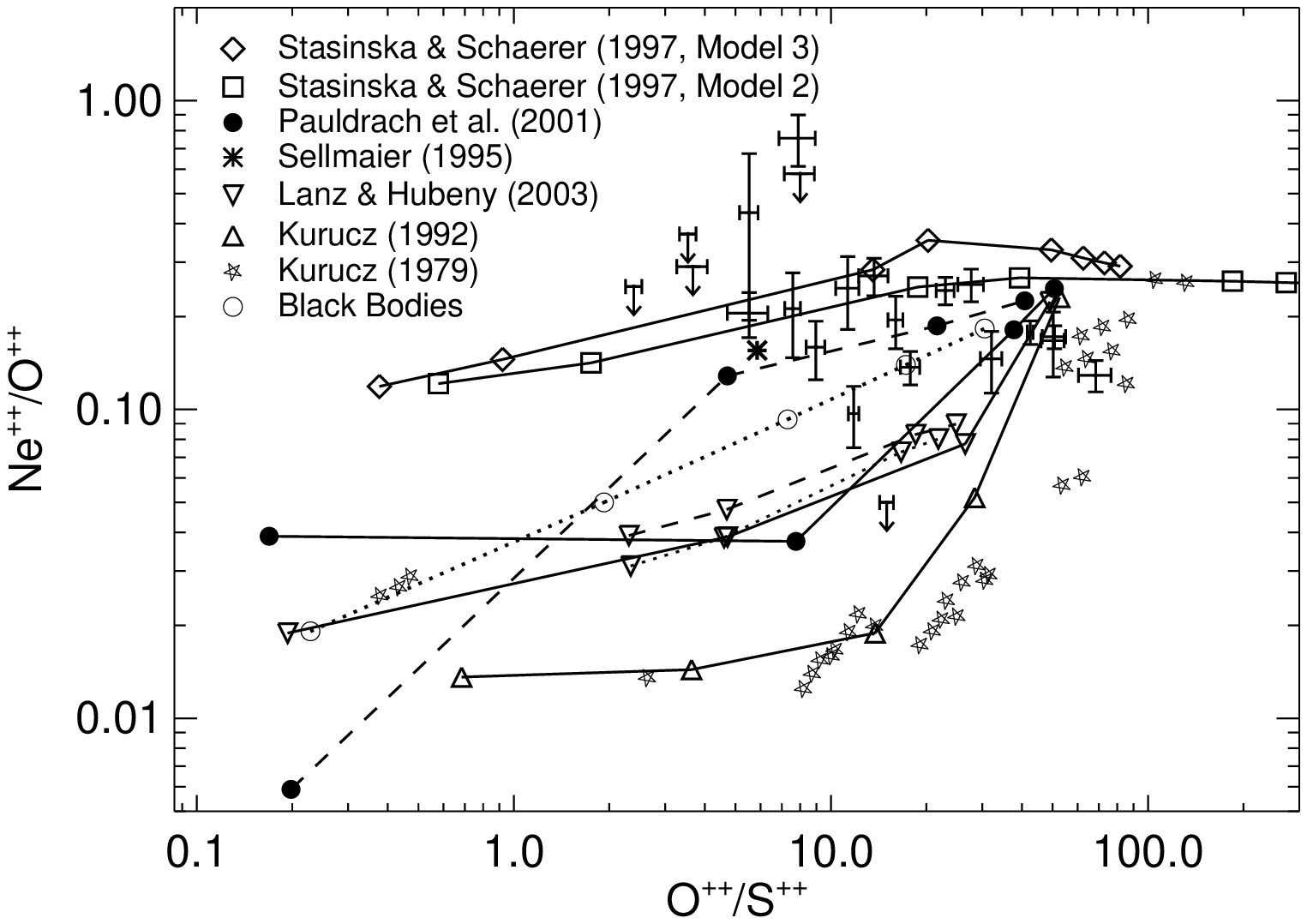}
\caption{Ratios of Ne$^{++}$/O$^{++}$ inferred from the observations of Simpson et al. (1995) plotted 
against their inferred O$^{++}$/S$^{++}$ ratios.
Ratios from the nebular models discussed in this paper are also plotted. 
The notations of 
Pauldrach et al. (2001), Sellmaier (private communication, 1995), 
Lanz \& Hubeny (2003), and Kurucz (1992) in the legend refer 
to the stellar atmospheres used in our NEBULA models
(same \teff\ as the models in Fig. 5; Sellmaier's atmosphere is a Supergiant with \teff\ = 35~kK).
The plotted Ne$^{++}$/O$^{++}$ and O$^{++}$/S$^{++}$ ratios for the models are equal to the calculated 
$<$Ne$^{++}$/Ne$>$/$<$O$^{++}$/O$>$ and $<$O$^{++}$/O$>$/$<$S$^{++}$/S$>$ ratios 
multiplied by assumed abundance ratios 
of Ne/O = 0.25 and O/S = 50, respectively
(Orion Nebula abundance ratios from Simpson et al. 1998).
For the \ion{H}{2} models calculated with Pauldrach et al. atmospheres, 
the solid line connects  
models with Dwarf atmospheres and the dashed line connects models with Supergiant atmospheres.
For the \ion{H}{2} region models calculated with Lanz \& Hubeny atmospheres, 
the solid line connects models with atmospheres with log $g = 4.0$
and the dotted and dashed lines connect models with atmospheres with log $g = 3.0$ to 3.5 and 
with Lyman continuum luminosities of $10^{49}$ and $10^{50}$ photons s$^{-1}$, respectively.
Models computed with atmospheres represented by black bodies from 20~kK to 40~kK
(every 5~kK) are also plotted (open circles connected by a dotted line).
}
\end{figure}

\clearpage


\begin{deluxetable}{lcccccccc} 
\tabletypesize{\scriptsize}
\tablewidth{0pt} 
\tablecaption{Observed Line and Continuum Fluxes. \label{tbl-1}} 
\tablehead{ 
\noalign{\centerline{\hskip3.0truein Line Flux\tablenotemark{a}\ \ ($10^{-18}$ W cm$^{-2}$) \hfill}} 
\noalign{\vskip2pt}
\noalign{\centerline{\hskip3.0truein Continuum Flux\tablenotemark{a}\ \ (Jy) \hfill}} 
\noalign{\vskip2pt}
\cline{2-9}
\colhead{} & \multicolumn{4}{c}{G333.6-0.2} & \colhead{} & 
\multicolumn{3}{c}{W43 (G30.8$-$0.0)} \\ 
\cline{2-5} \cline{7-9} \\ 
\colhead{Line} & \colhead{G333.6-ISO\tablenotemark{b}} & \colhead{G333.6-N1\tablenotemark{c}} & \colhead{G333.6-N2\tablenotemark{c}} & \colhead{G333.6-N3\tablenotemark{c}} & 
\colhead{} & \colhead{W43N\tablenotemark{c}} & \colhead{W43C\tablenotemark{c}} & 
\colhead{W43S\tablenotemark{c}}
}
\startdata 
\ [S III] 18.7 \micron\ & \nodata & 160.1$\pm$16.2 & 36.3$\pm$6.4\phn & \nodata & 
& SCREH & SCREH & SCREH \\ 
\ & \nodata & 1370$\pm$275& 290$\pm$280 & \nodata && SCREH & SCREH & SCREH \\
\ [S III] 33.5 \micron\ & \nodata & 189.1$\pm$3.6\phn & 65.7$\pm$4.7\phn & 38.9$\pm$3.1\phn 
& & SCREH & SCREH & SCREH \\ 
\ & \nodata & 7040$\pm$580 & 740$\pm$185 & 730$\pm$180 && SCREH & SCREH & SCREH \\
\ [O III] 51.8 \micron\ & 178.2$\pm$8.5 & 84.3$\pm$2.6\phn & 13.1$\pm$1.6 & 6.8$\pm$1.2 & 
& SCREH & SCREH & SCREH \\ 
\ & 54400 & 9910$\pm$320 & 1130$\pm$360 & 410$\pm$270 && SCREH & SCREH & SCREH \\
\ [N III] 57.3 \micron\ & 46.3$\pm$4.9 & 51.1$\pm$2.8\phn & 11.9$\pm$1.4 & \nodata & 
& SCREH & SCREH & SCREH \\ 
\ & 57630 & 10945$\pm$230\phn &1310$\pm$200 & \nodata && SCREH & SCREH & SCREH \\
\ [O III] 88.4 \micron\ & 39.9$\pm$3.9 & 37.7$\pm$0.8\phn & 14.5$\pm$0.9 & 7.5$\pm$0.7 & 
& SCREH & SCREH & SCREH \\ 
\ & 55100 & 9030$\pm$175\phn & 1060$\pm$270 & 630$\pm$90 && SCREH & SCREH & SCREH \\
\ [N II] 121.9 \micron\ & \nodata & 7.4$\pm$1.4\phn & 2.6$\pm$0.8 & \nodata & 
& 14.4$\pm$0.8 & 5.5$\pm$0.5 & 7.9$\pm$0.8 \\ 
\ & \nodata & 6790$\pm$120\phn & 575$\pm$230 & \nodata && 5270$\pm$420 & 1450$\pm$260 & 4690$\pm$380 \\ 
\ [N II] 121.9 \micron\tablenotemark{d} & \nodata & \nodata & \nodata & \nodata & 
& 4.9$\pm$0.5\tablenotemark{d} & 2.5$\pm$0.4\tablenotemark{d} & 2.6$\pm$0.6\tablenotemark{d} \\ 
\ & \nodata & \nodata & \nodata & \nodata && 2530$\pm$210\tablenotemark{d} & 800$\pm$200\tablenotemark{d} & 1600$\pm$300\tablenotemark{d} \\ 
\ [N II] 205.2 \micron\ & \nodata & \phn2.0$\pm$0.2\phn & 1.5$\pm$0.2 & 1.7$\pm$0.2 & 
& 1.52$\pm$0.26 & 0.46$\pm$0.14 & 0.56$\pm$0.13 \\ 
\ & \nodata & 2000$\pm$145 & 230$\pm$120 & 115$\pm$100 && 1300$\pm$70 & 490$\pm$50 & 1440$\pm$40 \\
\enddata 
\tablenotetext{a}{Flux and continuum errors are 1 $\sigma$ and are statistical only.}
\tablenotetext{b}{All ISO LWS data had entrance aperture $80''$.}
\tablenotetext{c}{All CGS data had entrance aperture $60''$ except as marked.}
\tablenotetext{d}{Entrance aperture $45''$.}
\end{deluxetable} 

\clearpage

\begin{deluxetable}{cccccccccc} 
\tabletypesize{\scriptsize}
\tablecolumns{10} 
\tablewidth{0pc} 
\tablenum{2}
\tablecaption{Electron Densities and Abundance Ratios} 
\tablehead{ 
\colhead{} & \multicolumn{5}{c}{G333.6-0.2} & \colhead{} & 
\multicolumn{3}{c}{W43} \\ 
\cline{2-6} \cline{8-10} \\ 
\colhead{} & \colhead{G333.6-ISO} & \colhead{G333.6-C\tablenotemark{a}} & \colhead{G333.6-N1} & \colhead{G333.6-N2} & 
\colhead{G333.6-N3} & \colhead{} & \colhead{W43N} & \colhead{W43C} & \colhead{W43S}
}
\startdata 
$\tau_{9.6}$ & 1.5 & 1.5 & 1.5 & 1.5 & 1.5 && 2.74 & 2.74 & 3.63 \\
$S_{\rm 5 GHz}\tablenotemark{b}$ & 40.75 & 26.65 & 5.35 & 1.10 & 0.35 && 5.91 & 2.25 & 3.11 \\
\sidehead{Electron Densities (cm$^{-3}$)}
$N_e$ (S III) & \nodata & 2400$\pm$350\tablenotemark{c} & 735$\pm$155\tablenotemark{d} & 335$\pm$150\tablenotemark{d} & \nodata &
& 340$\pm$120\tablenotemark{c} & 380$\pm$130\tablenotemark{c} & 280$\pm$160\tablenotemark{c} \\
$N_e$ (O III) & 2520$\pm$540\tablenotemark{e} & 4200$\pm$700\tablenotemark{c} & 730$\pm$50\tablenotemark{d} & 140$\pm$50\tablenotemark{d} & 140$\pm$70\tablenotemark{d} &
& 870$\pm$50\tablenotemark{c} & 730$\pm$60\tablenotemark{c} & 725$\pm$50\tablenotemark{c} \\
$N_e$ (N II) & \nodata & 330$^{+475}_{-185}$\tablenotemark{c} & 140$^{+65}_{-50}$\tablenotemark{d} & 37$^{+24}_{-21}$\tablenotemark{d} & \nodata &
& 5120$^{+\infty}_{-4200}$\tablenotemark{d} & $\infty$\tablenotemark{d} & $\infty$\tablenotemark{d} \\
$N_e$ (N II) & \nodata & \nodata & \nodata & \nodata & \nodata &
& 180$\pm$80\tablenotemark{f} & 650$^{+7940}_{-440}\tablenotemark{f}$ & 420$^{+835}_{-250}$\tablenotemark{f} \\
\sidehead{Ionic Abundance Ratios ($\times 10^{-6}$)}
S$^{++}$/H$^+$ & \nodata & 4.3$\pm$0.7 & 12.2$\pm$1.6 & 14.2$\pm$2.1 & 20.8$\pm$2.7 && 9.0$\pm$1.4 & 14.1$\pm$2.2 & 12.8$\pm$2.3 \\
O$^{++}$/H$^+$ & 38$\pm$5 & 71$\pm$9 & 80$\pm$8 & 50$\pm$6 & 82$\pm$12 && 130$\pm$14 & 275$\pm$31 & 227$\pm$25 \\
N$^{++}$/H$^+$ & 14$\pm$2 & 36$\pm$4 & 48$\pm$5 & 27$\pm$4 & \nodata && 68$\pm$8 & 127$\pm$14 & 117$\pm$13 \\
\sidehead{Ionic Abundance Ratios}
N$^{++}$/O$^{++}$ & 0.36$\pm$0.04 & 0.51$\pm$0.02 & 0.60$\pm$0.04 & 0.53$\pm$0.07 & \nodata && 0.52$\pm$0.03 & 0.46$\pm$0.02 & 0.52$\pm$0.02 \\
O$^{++}$/S$^{++}$ & \nodata & 16.7$\pm$2.2 & 6.53$\pm$0.55 & 3.56$\pm$0.57 & 3.95$\pm$0.52 && 14.5$\pm$1.2 & 19.5$\pm$1.9 & 17.8$\pm$1.8 \\
N$^{++}$/S$^{++}$ & \nodata & 8.47$\pm$0.70 & 3.90$\pm$0.29 & 1.89$\pm$0.29 & \nodata && 7.54$\pm$0.38 & 9.04$\pm$0.36 & 9.18$\pm$0.37 \\
N$^{+}$/S$^{++}$ & \nodata & 0.63$\pm$0.20 & 1.90$\pm$0.52 & 1.75$\pm$0.45 & 14.3$\pm$1.3 && 1.83$\pm$0.37\tablenotemark{g} & 3.98$\pm$1.67\tablenotemark{g} & 2.30$\pm$1.38\tablenotemark{g} \\
N$^{++}$/N$^{+}$ & \nodata & 13.3$\pm$4.0 & 2.06$\pm$0.45 & 1.08$\pm$0.27 & \nodata && 4.11$\pm$0.72\tablenotemark{g} & 2.27$\pm$0.84\tablenotemark{g} & 3.99$\pm$1.46\tablenotemark{g} \\
\sidehead{Column Density Ratios}
O$^{++}$/S$^{++}$ & \nodata & 9.55$\pm$0.43 & 6.56$\pm$0.22 & 8.60$\pm$0.87 & 3.95$\pm$0.52 && 5.64$\pm$0.22 & 10.2$\pm$0.5 & 6.89$\pm$0.47 \\
N$^{++}$/S$^{++}$ & \nodata & 4.85$\pm$0.23 & 3.92$\pm$0.24 & 4.56$\pm$0.67 & \nodata && 2.94$\pm$0.15 & 4.72$\pm$0.18 & 3.56$\pm$0.19 \\
N$^{+}$/S$^{++}$ & \nodata & 4.62$\pm$0.58 & 9.96$\pm$1.29 & 15.7$\pm$2.4 & 14.3$\pm$1.3 && 4.61$\pm$0.36\tablenotemark{h} & 3.29$\pm$0.37\tablenotemark{h} & 1.95$\pm$0.50\tablenotemark{h} \\
N$^{++}$/N$^{+}$ & \nodata & 1.05$\pm$0.13 & 0.39$\pm$0.08 & 0.29$\pm$0.09 & \nodata && 0.64$\pm$0.05\tablenotemark{h} & 1.44$\pm$0.14\tablenotemark{h} & 1.27$\pm$0.14\tablenotemark{h} \\

N$^{+}$/S$^{++}$ & \nodata & \nodata & \nodata & \nodata & \nodata && 3.39$\pm$0.41\tablenotemark{g} & 2.31$\pm$0.37\tablenotemark{g} & 1.55$\pm$0.34\tablenotemark{g} \\
N$^{++}$/N$^{+}$ & \nodata & \nodata & \nodata & \nodata & \nodata && 0.87$\pm$0.12\tablenotemark{g} & 2.04$\pm$0.34\tablenotemark{g} & 2.30$\pm$0.52\tablenotemark{g} \\

\sidehead{Adopted Abundance Ratios}
O$^{++}$/S$^{++}$ & \nodata & 13.1$\pm$3.6\tablenotemark{i} & 6.53$\pm$0.55 & 3.56$\pm$0.57 & 3.95$\pm$0.52 && 14.5$\pm$1.2 & 19.5$\pm$1.9 & 17.8$\pm$1.8 \\
N$^{++}$/S$^{++}$ & \nodata & 6.66$\pm$1.81\tablenotemark{i} & 3.90$\pm$0.29 & 1.89$\pm$0.29 & \nodata && 7.54$\pm$0.38 & 9.04$\pm$0.36 & 9.18$\pm$0.37 \\
N$^{+}$/S$^{++}$ & \nodata & 2.63$\pm$1.99\tablenotemark{i} & 5.93$\pm$4.03\tablenotemark{i} & 8.71$\pm$6.96\tablenotemark{i} & 14.3$\pm$6.0\tablenotemark{i} && 2.96$\pm$0.38\tablenotemark{j} & 3.02$\pm$0.80\tablenotemark{j} & 2.59$\pm$0.74\tablenotemark{j} \\
N$^{++}$/N$^{+}$ & \nodata & 1.05$\pm$0.13\tablenotemark{k} & 1.23$\pm$0.83\tablenotemark{i} & 0.69$\pm$0.39\tablenotemark{i} & \nodata && 1.43$\pm$0.30\tablenotemark{i} & 1.44$\pm$0.33\tablenotemark{i} & 1.89$\pm$0.53\tablenotemark{i} \\
(N$^{+}$+N$^{++}$)/S$^{++}$ & \nodata & 9.3$\pm$3.8 & 9.8$\pm$4.3 & 10.6$\pm$7.3 & $>$14$\pm$6 && 10.5$\pm$0.8 & 12.1$\pm$1.2 & 11.8$\pm$1.1 \\
\enddata 
\tablenotetext{a}{Center position, data from Colgan et al. (1993).}
\tablenotetext{b}{Radio flux in the aperture used for the FIR line measurement, 
scaled to the value appropriate for $\nu = 5$ GHz.}
\tablenotetext{c}{Electron density computed using the $45''$ aperture data.}
\tablenotetext{d}{Electron density computed using the $60''$ aperture data.}
\tablenotetext{e}{Electron density computed using the $80''$ aperture data.}
\tablenotetext{f}{Electron density computed using the $45''$ aperture data for
the 122 \micron\ line and reducing the 205 \micron\ line flux by the ratio of the aperture diameters.}
\tablenotetext{g}{Computed using the $45''$ aperture data for
the 122 \micron\ line and reducing the 205 \micron\ line flux by the ratio of the aperture diameters.}
\tablenotetext{h}{Computed using the $60''$ aperture data for
the 122 \micron\ line and reducing both the 122 \micron\ and the 205 \micron\ line fluxes by the ratio of the aperture diameters for comparison to the [N III] and [S III] lines.}
\tablenotetext{i}{Average of column density ratio and ionic abundance ratio; the error is increased for G333.6-N3 where $N_e$(N II) was not measured.}
\tablenotetext{j}{Average of ionic abundance ratios for all 122 \micron\ line measurements plus column density for the $45''$ aperture 122 \micron\ line measurements.}
\tablenotetext{k}{Column density ratio chosen because of high $N_e$ compared to $N_{crit}$.}
\end{deluxetable} 

\clearpage

\begin{deluxetable}{lccc} 
\tablecolumns{4} 
\tablewidth{0pc} 
\tablenum{3}
\tablecaption{Estimated Abundances} 
\tablehead{ 
\colhead{Element} & \colhead{G333.6-0.2} & \colhead{W43} & \colhead{Orion Nebula}
}
\startdata 
Estimated S/H\tablenotemark{a} & 13.4$\pm0.8 \times10^{-6}$ & 12.1$\pm1.1 \times10^{-6}$ & $7.0 \times10^{-6}$\ \tablenotemark{b} \\ 
Estimated N/S\tablenotemark{a} & 9.3$\pm$2.1 & 10.3$\pm$0.4 & 9.71\tablenotemark{b} \\ 
Estimated N/H\tablenotemark{c} & 12.5$\times10^{-5}$ & 12.5$\times10^{-5}$ & 6.8$\times10^{-5}$~\tablenotemark{d} \\ 
Estimated Ne/H & 3.4$\times10^{-4}$\ \tablenotemark{e} & 1.1$\times10^{-4}$\ \tablenotemark{f} & 1.0$\times10^{-4}$\ \tablenotemark{b} \\
Estimated O/H & 3 to 4$\times10^{-4}\ \tablenotemark{g}$ & 3 to 8$\times10^{-4}$\ \tablenotemark{g} & 4.0$\times10^{-4}$\ \tablenotemark{d} \\
\enddata 
\tablenotetext{a}{Assumed fractional ionization $<$S$^{++}$/S$> = 0.9$.}
\tablenotetext{b}{Simpson et al. (1998) revised for new collisional excitation cross sections (see text).}
\tablenotetext{c}{N/H = N/S $\times$ S/H.}
\tablenotetext{d}{Rubin et al. (1991).}
\tablenotetext{e}{Estimated from [Ne II] 12.8 \micron\ measurements of Fujiyoshi et al. (1998) and radio continuum measurements of Fujiyoshi (1999).}
\tablenotetext{f}{SCREH.}
\tablenotetext{g}{O/H = O$^{++}$/H$^+ \times ICF_{\rm Model}$.}
\end{deluxetable} 

\clearpage

\begin{deluxetable}{lccccccccccccc} 
\tabletypesize{\scriptsize}
\tablecolumns{14} 
\tablewidth{0pc} 
\tablenum{4}
\tablecaption{G333.6-0.2 - Comparison of Models to Observations} 
\tablehead{ 
\colhead{} & \multicolumn{4}{c}{Observed Fluxes and Ratios\tablenotemark{a}} & 
\colhead{} & \multicolumn{4}{c}{Model for G333.6-0.2 Core\tablenotemark{a,b}} &
\colhead{} & \multicolumn{3}{c}{Model for North Positions\tablenotemark{a,b}} \\ 
\cline{2-5} \cline{7-10} \cline{12-14} \\ 
\colhead{Line Flux or Ratio} & \colhead{Center} & \colhead{N1} & \colhead{N2} & \colhead{N3} &
\colhead{} & \colhead{Center} & \colhead{N1} & \colhead{N2} & \colhead{N3} &
\colhead{} & \colhead{N1} & \colhead{N2} & \colhead{N3} \\
}
\startdata
\ [O III] F$_{52}$ \micron & 161\tablenotemark{c} & 92 & 14 & 7 && 350 & 0.0005 & 0 & 0 && 117 & 1 & 0 \\
\ [O III] F$_{88}$ \micron & 27\tablenotemark{c} & 39 & 15 & 8 && 54 & 0.0001 & 0 & 0 && 67 & 1 & 0 \\
\ [O III] F$_{52}$/F$_{88}$ & 6.06 & 2.37 & 0.96 & 0.96 && 6.50 & 4.12 & \nodata & \nodata && 1.75 & 1.61 & \nodata \\
\ [S III] F$_{19}$ \micron & 412\tablenotemark{c} & 270 & 61 & \nodata && 869 & 273 & 50 & 14 && 154 & 52 & 16 \\
\ [S III] F$_{33}$ \micron & 158\tablenotemark{c} & 228 & 79 & 47 && 351 & 243 & 72 & 27 && 192 & 85 & 33 \\
\ [S III] F$_{19}$/F$_{33}$ & 2.61 & 1.19 & 0.78 & \nodata && 2.48 & 1.13 & 0.70 & 0.52 && 0.80 & 0.61 & 0.48 \\
\ [N II] F$_{122}$ \micron & 2.5\tablenotemark{c} & 7.4 & 2.6 & \nodata && 9.2 & 13.7 & 5.6 & 2.8 && 6.8 & 6.9 & 3.5 \\
\ [N II] F$_{205}$ \micron & 0.44\tablenotemark{c} & 1.97 & 1.46 & 1.66 && 1.5 & 2.4 & 1.4 & 0.9 && 1.9 & 2.0 & 1.4 \\
\ [N II] F$_{122}$/F$_{205}$ & 5.7 & 3.7 & 1.8 & \nodata && 6.2 & 5.8 & 4.1 & 3.0 && 3.7 & 3.4 & 2.5 \\
\ [N III] F$_{57}$ \micron & 52\tablenotemark{c} & 55 & 13 & \nodata && 115 & 0.006 & 0 & 0 && 62 & 1.9 & 0 \\
\ [N III] F$_{57}$/[N II] F$_{122}$ & 21.0 & 7.4 & 5.0 & \nodata && 12.6 & 0.0004 & 0 & 0 && 9.1 & 0.28 & 0 \\
\ [N III] F$_{57}$/[O III] F$_{52}$ & 0.32 & 0.60 & 0.89 & \nodata && 0.33 & 12 & \nodata & \nodata && 0.53 & 1.49 & \nodata \\
\ [Ne III] F$_{36}$ \micron\tablenotemark{c} & 17 & \nodata & \nodata & \nodata && 8.5 & \nodata & \nodata & \nodata && \nodata & \nodata & \nodata \\
\ [Ne II] F$_{12.8}$ \micron\tablenotemark{d} & 1300 & \nodata & \nodata & \nodata && 1793 & \nodata & \nodata & \nodata && \nodata & \nodata & \nodata \\
\ [S IV] F$_{10.5}$ \micron\tablenotemark{e} & 100 & \nodata & \nodata & \nodata && 95 & \nodata & \nodata & \nodata && \nodata & \nodata & \nodata \\
\ [Ar III] F$_{9.0}$ \micron\tablenotemark{f} & 350 & \nodata & \nodata & \nodata && 251 &\nodata & \nodata & \nodata && \nodata & \nodata & \nodata \\
\ [Ar III] F$_{9.0}$ \micron\tablenotemark{g} & 83 & \nodata & \nodata & \nodata && 150 & \nodata & \nodata & \nodata && \nodata & \nodata & \nodata \\
\ [Ar II] F$_{7.0}$ \micron\tablenotemark{g} & 74 & \nodata & \nodata & \nodata && 13 & \nodata & \nodata & \nodata && \nodata & \nodata & \nodata \\
\ He$^+$/H$^+$ RRLs\tablenotemark{h} & 0.045 & \nodata & \nodata & \nodata && 0.044 & \nodata & \nodata & \nodata && \nodata & \nodata & \nodata \\
\ [O III] F$_{52}$ \micron\tablenotemark{i} & 194 & \nodata & \nodata & \nodata && 359 & \nodata & \nodata & \nodata && \nodata & \nodata & \nodata \\
\ [O III] F$_{88}$ \micron\tablenotemark{i} & 41 & \nodata & \nodata & \nodata && 56 & \nodata & \nodata & \nodata && \nodata & \nodata & \nodata \\
\ [N III] F$_{57}$ \micron\tablenotemark{i} & 50 & \nodata & \nodata & \nodata && 144 & \nodata & \nodata & \nodata && \nodata & \nodata & \nodata \\
\enddata
\tablenotetext{a}{Fluxes $\times 10^{-18}$ W cm$^{-2}$, corrected for extinction.}
\tablenotetext{b}{Model Abundances ($\times 10^{-6}$): C/H = 400, N/H = 125, O/H = 500, Ne/H = 340, S/H = 13.4, Ar/H = 4.5. }
\tablenotetext{c}{$0.75'$ Aperture, Colgan et al. (1993).}
\tablenotetext{d}{$0.75'$ Aperture, Geballe et al. (1981).}
\tablenotetext{e}{$0.50'$ Aperture, Geballe et al. (1981).}
\tablenotetext{f}{$0.67'$ Aperture, Geballe et al. (1981).}
\tablenotetext{g}{$0.35'$ Aperture, Cohen et al. (1989), Simpson et al. (1995b).}
\tablenotetext{h}{Radio recombination lines measured in $2.3'$ aperture, McGee \& Newton (1981).}
\tablenotetext{i}{$80''$ Aperture for ISO LWS.}
\end{deluxetable} 

\clearpage

\begin{deluxetable}{lcccccccl} 
\tabletypesize{\scriptsize}
\tablecolumns{9} 
\tablewidth{0pc} 
\tablenum{5}
\tablecaption{W43 - Comparison of Models to Observations} 
\tablehead{ 
\colhead{} & \multicolumn{3}{c}{Observed Fluxes and Ratios\tablenotemark{a}} & 
\colhead{} & \multicolumn{3}{c}{Model Fluxes\tablenotemark{b} and Ratios} \\
\cline{2-4} \cline{6-8}  \\ 
\colhead{Line Flux or Ratio} & \colhead{W43N} & \colhead{W43C} & \colhead{W43S} & \colhead{} & 
\colhead{Model D\tablenotemark{c}} & \colhead{Model S\tablenotemark{c}} & \colhead{Model T\tablenotemark{c}} & \colhead{Comments \hfill} }
\startdata 
\ $S_{\rm 5GHz}$ & 5.91 & 2.25 & 3.11 && 3.50 & 3.50 & 3.50 &\\
\ He$^+$/H$^+$ RRLs && 0.076\tablenotemark{d} &&& 0.075 & 0.073 & 0.075 & Good $T_{\rm eff}$ \\
\ [S III] F$_{19}$ \micron & 202 & 122 & 149 && 155 & 161 & 163 &\\
\ [S III] F$_{33}$ \micron & 261 & 149 & 208 && 149 & 156 & 160 &\\
\ [S III] 19/33 \micron\ & 0.78 & 0.82 & 0.72 && 1.04 & 1.03 & 1.02 & Model $N_e$ too high \\
\ [O III] F$_{52}$ \micron & 153 & 130 & 148 && 205 & 143 & 81 &\\
\ [O III] F$_{88}$ \micron & 59 & 56 & 64 && 89 & 62 & 35 &\\
\ [O III] 52/88 \micron\ & 2.6 & 2.3 & 2.3 && 2.30 & 2.33 & 2.33 & Good $N_e$ \\
\ [N II] F$_{122}$ \micron\tablenotemark{e} & 11 & 4.1 & 5.9 && 4.2 & 4.7 & 4.2 &\\
\ [N II] F$_{122}$ \micron & 4.9 & 2.5 & 2.6 && 4.2 & 4.7 & 4.2 &\\
\ [N II] F$_{205}$ \micron\tablenotemark{e} & 1.14 & 0.35 & 0.42 && 0.66 & 0.75 & 0.68 &\\
\ [N II] 122/205 \micron\ & 4.3 & 7.1 & 6.2 && 6.29 & 6.28 & 6.19 & Good $N_e$ \\
\ [N III] F$_{57}$ \micron & 76 & 61 & 77 && 61 & 58 & 65 &\\
\ [N III] F$_{57}$/[O III] F$_{52}$ & 0.50 & 0.47 & 0.52 && 0.30 & 0.41 & 0.80 &\\
\ [N III] F$_{57}$/[N II] F$_{122}$\tablenotemark{e} & 7 & 15 & 13 && 15 & 12 & 15 &\\
\ [N III] F$_{57}$/[N II] F$_{122}$ & 15 & 23 & 30 && 15 & 12 & 15 &\\
\ S$^{++}$/H$^+$ & 9.0 & 14.1 & 12.8 && \nodata & \nodata & \nodata & Times $10^{-6}$ \\
\ S$^{++}$/S & \nodata & \nodata & \nodata && 0.919 & 0.953 & 0.965 & Implies S/H = 11.8, 11.4, or 11.3$\times 10^{-6}$ \\
\ O$^{++}$/S$^{++}$ & 14.5 & 19.5 & 17.8 && 0.583\tablenotemark{f} & 0.391\tablenotemark{f} & 0.221\tablenotemark{f} & Implies O/H = 3.3, 4.8, or 8.3$\times 10^{-4}$\ \tablenotemark{g} \\
\ N$^{++}$/O$^{++}$ & 0.521 & 0.463 & 0.516 && 1.18 & 1.58 & 3.05 & Implies N/O = 0.42, 0.32, or 0.16\tablenotemark{h} \\
\ N$^{++}$/N$^{+}$ & 0.6--2.5 & 1.4--2.2 & 1.3--3.2 && 1.72 & 1.45 & 1.86 & Compare uniform He$^+$/H$^+$ \\
\ N$^{+}$/S$^{++}$ & 3.0 & 3.0 & 2.6 && 0.400 & 0.428 & 0.363 & Implies N/S = 7.25, 6.78, or 7.99\tablenotemark{i} \\
\ N$^{++}$/S$^{++}$ & 7.5 & 9.0 & 9.2 && 0.686 & 0.620 & 0.673 & Implies N/S = 12.55, 13.89, or 12.79\tablenotemark{i} \\
\enddata 
\tablenotetext{a}{Fluxes $\times 10^{-18}$ W cm$^{-2}$, corrected for extinction. Beam size is $45''$.}
\tablenotetext{b}{Model fluxes are scaled from the total model 5~GHz flux ($\simeq 92$ Jy) to 3.5 Jy since the KAO beams are much smaller than the total diameter of W43.}
\tablenotetext{c}{ 
The model abundance ratios ($\times 10^{-6}$) are
C/H = 400, N/H = 125, O/H = 500, Ne/H = 110, S/H = 12.1, and Ar/H = 4.5. }
\tablenotetext{d}{Weighted average of measurements of Churchwell et al. (1978), Lichten,
Rodriguez, \& Chaisson (1979), Lockman \& Brown (1982), Peimbert et al. (1992).}
\tablenotetext{e}{Flux measured in a $60''$ beam and scaled to $45''$.}

\tablenotetext{f}{Model O$^{++}$/S$^{++}$ is the density weighted ratio of $<$O$^{++}$/O$>$/$<$S$^{++}$/S$>$, etc. (SCREH). }
\tablenotetext{g}{O/H = O/S $\times$ S/H.}
\tablenotetext{h}{N/H = 1.4, 1.5, or 1.4 $\times 10^{-4}$ if O/H = 3.3, 4.8, or 8.3 $\times 10^{-4}$.}
\tablenotetext{i}{The average N/S = 9.9, 10.3, or 10.4, N/H = 1.17, 1.18, or 1.17 $\times 10^{-4}$, and O/H = 2.8, 3.7, or 7.2 $\times 10^{-4}$.}
\end{deluxetable} 



\newpage

\begin{thebibliography}{}
\bibitem[Afflerbach et al.(1997)]{acw97}Afflerbach, A., Churchwell, E., \& Werner, M. W. 1997, \apj, 478, 190

\bibitem[Anders \& Grevesse (1989)]{ag89}Anders, E., \& Grevesse, N. 1989, \gca, 53, 197

\bibitem[Balser et al. (1995)]{bbw95}Balser, D. S., Bania, T. M., Rood, R. T., \& Wilson, T. L. 1995, \apjs, 100, 371
 
\bibitem[Blum et al. (2002)]{bcd02}Blum, R. D., Conti, P. S., Damineli, A., \& Figuer\^edo, E. 2002, in 
ASP Conf. Ser. 267, Hot Star Workshop III: The Earliest Phases of Massive Star Birth, 
ed. P. A. Crowther (San Francisco: ASP), 283

\bibitem[Blum et al. (1999)]{bdc99}Blum, R. D., Damineli, A., \& Conti, P. S. 1999, \aj, 117, 1392

\bibitem[Blum \& Pradhan (1992)]{bp92}Blum, R. D., \& Pradhan, A. K. 1992, \apjs, 80, 425

\bibitem[Caswell \& Haynes (1987)]{ch87}Caswell, J. L., \& Haynes, R. F. 1987, \aap, 171, 261

\bibitem[Chiappini et al. (2001)]{chi01}Chiappini, C., Matteucci, F., \& Romano, D. 2001, \apj, 554, 1044

\bibitem[Chiappini et al. (2003)]{chi03}Chiappini, C., Romano, D., \& Matteucci, F. 2003, \mnras, 339, 63

\bibitem[Churchwell et al. (1978)]{csm78}Churchwell, E., Smith, L. F., Mathis, J., Mezger, P. G., \& Huchtmeier, W. 1978, \aap, 70, 719

\bibitem[Churms et al. (1974)]{cfg74}Churms, J., Feast, M. W., Glass, I. S., Harding, G. A., Lloyd Evans, T., \& Martin, W. L. 1974, \mnras, 169, 39P 

\bibitem[Cohen et al. (1989)]{coh89}Cohen, M., et al. 1989, \apj, 341, 246

\bibitem[Colgan et al. (1993)]{col93}Colgan, S. W. J., Haas, M. R., Erickson, E. F., Rubin, R. H., Simpson, J. P., \& Russell, R. W. 1993, \apj, 413, 237

\bibitem[Colgan et al. (1991)]{col91}Colgan, S. W. J., Simpson, J. P., Rubin, R. H., Erickson, E. F., Haas, M. R., \& Wolf, J. 1991, \apj, 366, 172

\bibitem[Condon et al. (1998)]{con98}Condon, J. J., Cotton, W. D., Greisen, E. W., Yin, Q. F., Perley, R. A., Taylor, G. B., \& Broderick, J. J. 1998, \aj, 115, 1693

\bibitem[Cotera \& Simpson (1997)]{cs97}Cotera, A., \& Simpson, J. 1997, \baas, 29, 1397

\bibitem[Erickson et al. (1985)]{efe85}Erickson, E. F., et al. 1985, Infrared Phys., 25, 513

\bibitem[Esteban et al. (1998)]{ept98}Esteban, C., Peimbert, M., Torres-Peimbert, S., \& Escalante, V. 1998, \mnras, 295, 401

\bibitem[Forster et al. (1987)]{for87}Forster, J. R., Whiteoak, J. B., Kesteven, M. J., Manchester, R. N., Rayner, P. T., Ables, J. G., \& Peters, W. L. 1987, \mnras, 226, 173

\bibitem[Fujiyoshi (1999)]{fuj99}Fujiyoshi, T. 1999, in Star Formation 1999, ed. T. Nakamoto (Nobeyama, Japan: Nobeyama Radio Observatory), 401

\bibitem[Fujiyoshi et al. (1998)]{fuj98}Fujiyoshi, T., Smith, C. H., Moore, T. J. T., Aitken, D. K., Roche, P. F., \& Quinn, D. E. 
1998, \mnras, 296, 225

\bibitem[Fujiyoshi et al. (2001)]{fuj01}Fujiyoshi, T., Smith, C. H., Wright, C. M., Moore, T. J. T., Aitken, D. K., \& Roche, P. F. 
2001, \mnras, 327, 233

\bibitem[Galav\'{\i}s et al. (1995)]{Ga95} Galav\'{\i}s, M. E., Mendoza, C., \& Zeippen, C. J. 1995, \aaps, 111, 347

\bibitem[Galav\'{\i}s et al. (1998)]{Ga98} Galav\'{\i}s, M. E., Mendoza, C., \& Zeippen, C. J. 1998, \aaps, 133, 245

\bibitem[Geballe et al. (1981)]{geb81}Geballe, T. R., Wamsteker, W., Danks, A. C., Lacy, J. H., \& Beck, S. C. 1981, \apj, 247, 130

\bibitem[Giveon et al. (2002)]{giv02}Giveon, U., Sternberg, A., Lutz, D., Feuchtgruber, H., \& Pauldrach, A. W. A. 2002, \apj, 566, 880

\bibitem[Henry \& Worthy (1999)]{hw99}Henry, R. B. C., \& Worthey, G. 1999, \pasp, 111, 919

\bibitem[Holweger (2001)]{hol01}Holweger, H. 2001, in AIP Conf. Proc. 598, Solar and Galactic Composition: 
A Joint SOHO/ACE Workshop, ed. R. F. Wimmer-Schweingruber (New York: AIP), 23

\bibitem[Hou et al. (2000)]{hou00}Hou, J. L., Prantzos, N., \& Boissier, S. 2000, \aap, 362, 921

\bibitem[Hyland et al. (1980)]{hyl80}Hyland, A. R., McGregor, P. J., Robinson, G., Thomas, J. A., Becklin, E. E., Gatley, I., \& Werner, M. W. 1980, \apj, 241, 709
 
\bibitem[Kennicutt et al. (2003)]{kbg03}Kennicutt, R. C. Jr., Bresolin, F., \& Garnett, D. R. 
2003, \apj, 591, 801
 
\bibitem[Kingdon \& Ferland (1996)]{kin96}Kingdon, J. B., \& Ferland, G. J. 1996, \apjs, 106, 205

\bibitem[Kisielius \& Storey (2002)]{kis02}Kisielius, R., \& Storey, P. J. 2002, \aap, 387, 1135

\bibitem[Kisielius et al. (1998)]{kis98}Kisielius, R., Storey, P. J., Davey, A. R., \& Neale, L. T. 1998, \aaps, 133, 257

\bibitem[Kurucz (1979)]{kur79}Kurucz, R. L. 1979, \apjs, 40, 1

\bibitem[Kurucz (1992)]{kur92}Kurucz, R. L. 1992, in IAU Symp. 149, Stellar Population of Galaxies, 
ed. B. Barbuy \& A. Renzini (Dordrecht: Kluwer), 225

\bibitem[Lanz \& Hubeny (2003)]{lh03}Lanz, T., \& Hubeny, I. 2003, \apjs, 146, 417

\bibitem[Lennon \& Burke (1994)]{lb94}Lennon, D. J., \& Burke, V. M. 1994, \aaps, 103, 273
 
\bibitem[Lester et al. (1985)]{ldw85}Lester, D. F., Dinerstein, H. L., Werner, M. W., Harvey, P. M., Evans, N. J. II, \& Brown, R. L. 1985, \apj, 296, 565

\bibitem[Li \& Draine (2001)]{ld01}Li, A. \& Draine, B. T. 2001, \apj, 554, 778

\bibitem[Lichten et al. (1979)]{lrc79}Lichten, S. M., Rodriguez, L. F., \& Chaisson, E. J. 1979, \apj, 229, 524

\bibitem[Liszt (1995)]{l95}Liszt, H. S. 1995, \aj, 109, 1204

\bibitem[Lockman \& Brown (1982)]{lb82}Lockman, F. J., \& Brown, R. L. 1982, \apj, 259, 595

\bibitem[Mart\'{\i}n-Hern\'{a}ndez et al. (2002a)]{mh02a}Mart\'{\i}n-Hern\'{a}ndez, N. L., et al. 2002a, \aap, 381, 606

\bibitem[Mart\'{\i}n-Hern\'{a}ndez et al. (2003)]{mh03}Mart\'{\i}n-Hern\'{a}ndez, N. L., van der Hulst, J. M., \& Tielens, A. G. G. M. 2003, \aap, 407, 957

\bibitem[Mart\'{\i}n-Hern\'{a}ndez et al. (2002b)]{mh02b}Mart\'{\i}n-Hern\'{a}ndez, N. L., Vermeij, R., Tielens, A. G. G. M., van der Hulst, J. M., \& Peeters, E. 2002b, \aap, 389, 286

\bibitem[McGee \& Newton (1981)]{mgn81}McGee, R. X., \& Newton, L. M. 1981, \mnras, 196, 889
 
\bibitem[McGee et al. (1979)]{mgn79}McGee, R. X., Newton, L. M., \& Butler, P. W. 1979, \mnras, 189, 413

\bibitem[McLaughlin \& Bell (2000)]{mlb00}McLaughlin, B. M., \& Bell, K. L. 2000, J.Phys.B, 33, 597

\bibitem[Mezger \& Henderson (19667)]{mh67}Mezger, P. G., \& Henderson, A. P. 1967, \apj, 147, 471

\bibitem[Morisset (2004)]{mor04}Morisset, C. 2004, \apj, 601, 858

\bibitem[Morisset et al. (2004)]{msbm04}Morisset, C., Schaerer, D., Bouret, J.-C., \& Martins, F. 2004, \aap, 415, 577

\bibitem[Morisset et al. (2002)]{mor02}Morisset, C., et al. 2002, \aap, 386, 558
 
\bibitem[Motte et al. (2003)]{msl03}Motte, F., Schilke, P., \& Lis, D. C. 2003, \apj, 582, 277

\bibitem[Nahar (1995)]{nah95}Nahar, S. N. 1995, \apjs, 101, 423

\bibitem[Nahar (1999)]{nah99}Nahar, S. N. 1999, \apjs, 120, 131

\bibitem[Nahar (2000)]{nah00}Nahar, S. N. 2000, \apjs, 126, 537

\bibitem[Nahar \& Pradhan (1997)]{nah97}Nahar, S. N., \& Pradhan, A. K. 1997, \apjs, 111, 339

\bibitem[Osterbrock (1989)]{ost89}Osterbrock, D. E. 1989, Astrophysics of Gaseous Nebulae \& Active Galactic Nuclei (Mill Valley, CA: University Science Books)

\bibitem[Pagel (2001)]{pag01}Pagel, B. E. J. 2001, \pasp, 113, 137

\bibitem[Pauldrach et al. (2001)]{phl01}Pauldrach, A. W. A., Hoffmann, T. L., \& Lennon, M. 2001, \aap, 375, 161

\bibitem[Peimbert et al. (1992)]{pei92}Peimbert, M., Rodr\'{\i}guez, L. F., Bania, T. M., Rood, R. T., \& Wilson, T. L. 1992, \apj, 395, 484

\bibitem[Pelen \& Berrington (1995)]{Pe95}Pelen, J., \& Berrington, K. A. 1995, \aaps, 110, 209

\bibitem[Price et al. (2001)]{pri01}Price, S. D., Egan, M. P., Carey, S. J., Mizuno, D. R., \& Kuchar, T. A. 2001, \aj, 121, 2819

\bibitem[Rank et al. (1978)]{rdl78}Rank, D. M., Dinerstein, H. L., Lester, D. F., Bregman, J. D., Aitken, D. K., \& Jones, B. 1978, \mnras, 185, 179

\bibitem[Reifenstein et al. (1970)]{rei70}Reifenstein, E. C. III, Wilson, T. L., Burke, B. F., Mezger, P. G., \& Altenhoff, W. J. 1970, \aap, 4, 357
 
\bibitem[Retallack \& Goss (1980)]{ret80}Retallack, D. S., \& Goss, W. M. 1980, \mnras, 193, 261

\bibitem[Rodr\'iguez \& Rubin (2004)]{rod04}Rodr\'{\i}guez, M., \& Rubin, R. H. 2004,  
in IAU Symposium No. 217, ASP Conf. Ser., Recycling Intergalactic and Interstellar Matter,
ed. P.-A. Duc, J. Braine, \& E. Brinks (San Francisco: ASP), in press (astro-ph/0312246)

\bibitem[Rolleston et al. (2000)]{rol01}Rolleston, W. R. J., Smartt, S. J., Dufton, P. L., \& Ryans, R. S. I. 2000, \aap, 363, 537

\bibitem[Rubin (1968a)]{rub68a}Rubin, R. H. 1968a, \apj, 153, 761
 
\bibitem[Rubin (1968b)]{rub68b}Rubin, R. H. 1968b, \apj, 154, 391
 
\bibitem[Rubin (1985)]{rub85}Rubin, R. H. 1985, \apjs, 57, 349
 
\bibitem[Rubin (1989)]{rub89}Rubin, R. H.  1989, \apjs, 69, 897
 
\bibitem[Rubin et al. (1983)]{rhe83}Rubin, R. H., Hollenbach, D. J., \& Erickson, E. F. 1983, \apj, 265, 239

\bibitem[Rubin et al. 1995]{rky95}Rubin, R. H., Kunze, D., \& Yamamoto, T. 1995, 
in ASP Conf. Ser. 78, Astrophysical Applications of Powerful New Atomic Databases, 
ed. S. J. Adelman \& W. L. Wiese (San Francisco: ASP), 479  

\bibitem[Rubin et al. (1991)]{rsh91}Rubin, R. H., Simpson, J. P., Haas, M. R., \& Erickson, E. F. 1991, \apj, 374, 564
 
\bibitem[Rubin et al. (1994)]{rsl94}Rubin, R. H., Simpson, J. P., Lord, S. D., Colgan, S. W. J., Erickson, E. F., \& Haas, M. R. 1994, \apj, 420, 772

\bibitem[Rudolph et al. (1997)]{rud97}Rudolph, A. L., Simpson, J. P., Haas, M. R., Erickson, E. F., \& Fich, M. 1997, \apj, 489, 94

\bibitem[Saraph \& Storey (1999)]{ss99}Saraph, H. E., \& Storey, P. J. 1999, \aaps, 134, 369

\bibitem[Saraph \& Tully (1994)]{Sa94}Saraph, H. E., \& Tully, J. A. 1994, \aaps, 107, 29

\bibitem[Schaerer \& de Koter (1997)]{sdk97}Schaerer, D., \& de Koter, A. 1997, \aap, 322, 598

\bibitem[Schraml \& Mezger (1969)]{sm69}Schraml, J., \& Mezger, P. G. 1969, \apj, 156, 269

\bibitem[Seaton et al. (1992)]{sea92}Seaton, M. J., Zeippen, C. J., Tully, J. A., Pradhan, A. K., Mendoza, C., Hibbert, A., \& Berrington, K. A. 1992, Rev. Mexicana Astron. Astrofis., 23, 19

\bibitem[Sellmaier et al. (1996)]{sel96}Sellmaier, F., Yamamoto, T., Pauldrach, A. W. A., \& Rubin, R. H. 1996, \aap, 305, L37
 
\bibitem[SBDLRWW]{Simpson et al. 1995b}Simpson, J. P., Bregman, J. D., Dinerstein, H. L., 
Lester, D. F., Rank, D. M., Witteborn, F. C., \& Wooden, D. H. 1995b, in ASP Conf. Ser. 73, 
Airborne Astronomy Symposium on the Galactic Ecosystem: From Gas to Stars to Dust,
ed. M. R. Haas, J. A. Davidson, \& E. F. Erickson (San Francisco: ASP), 105

\bibitem[SCREH]{screh}Simpson, J. P., Colgan, S. W. J., Rubin, R. H., Erickson, E. F.,  \& Haas, M. R. 1995a, \apj, 444, 721 (SCREH)
 
\bibitem[Simpson \& Rubin (1990)]{sr90}Simpson, J. P., \& Rubin, R. H. 1990, \apj, 354, 165
 
\bibitem[Simpson et al. (1986)]{sreh}Simpson, J. P., Rubin, R. H., Erickson, E. F., \& Haas, M. R. 1986, \apj, 311, 895

\bibitem[Simpson et al. (1998)]{swpc}Simpson, J. P., Witteborn, F. C., Price, S. D., \& Cohen, M. 1998, \apj, 508, 268

\bibitem[Smith et al. (2002)]{snc02}Smith, L. J., Norris, R. P. F., \& Crowther, P. A. 2002, \mnras, 337, 1309

\bibitem[Stasi\'nska \& Schaerer (1997)]{ss97}Stasi\'nska, G., \& Schaerer, D. 1997, \aap, 322, 615

\bibitem[Subrahmanyan \& Goss (1996)]{sg96}Subrahmanyan, R., \& Goss, W. M. 1996, \mnras, 281, 239

\bibitem[Tayal (2000)]{tay00}Tayal, S. S. 2000, \apj, 530, 1091

\bibitem[Tayal \& Gupta (1999)]{tay99}Tayal, S. S., \& Gupta, G. P. 1999, \apj, 526, 544
 
\bibitem[Vall\'ee (2002)]{val02}Vall\'ee, J. P. 2002, \apj, 566, 261

\bibitem[Verner et al. (1996)]{vfky96}Verner, D. A., Ferland, G. J., Korista, K. T., \& Yakovlev, D. G. 1996, \apj, 465, 487

\bibitem[Wilson et al. (1970)]{wil70}Wilson, T. L., Mezger, P. G., Gardner, F. F., \& Milne, D. K. 1970, \aap, 6, 364

\bibitem[Wink et al. (1983)]{wwb83}Wink, J. E., Wilson, T. L., \& Bieging, J. H. 1983, \aap, 127, 211
\end{thebibliography}
\end{document}